\documentclass[reprint, aps, amsmath, amssymb, superscriptaddress, nofootinbib]{revtex4-2}
\usepackage{amsmath}
\usepackage{siunitx}
\usepackage{verbatim}
\usepackage{bm}
\usepackage{ulem}
\usepackage{helvet}
\usepackage[bookmarks, colorlinks=false]{hyperref}
\usepackage{mathtools}
\hypersetup{colorlinks=true, allcolors=blue}
\usepackage{verbatim}
\usepackage{csquotes}
\usepackage{xcolor}
\usepackage{multirow}
\usepackage{subfigure}

\usepackage{graphics}
\usepackage{graphicx}

\renewcommand{\emph}[1]{{\textit{\textcolor{blue}{#1}}}}

\newcommand{\minifrac}[2]{\ensuremath{^{#1}\!/\!_{#2}}}

\begin{document}

\title{Multiscale modelling of diffusion and retention of hydrogen in multi-occupancy traps in irradiated bcc metals}

\author{Daniel R. Mason}
	\email{daniel.mason@ukaea.uk}
	\affiliation{UK Atomic Energy Authority, Culham Campus, Abingdon,  Oxfordshire OX14 3DB, United Kingdom}

\author{Sanjeet Kaur}
	%\email{sanjeet.kaur@ukaea.uk}
	\affiliation{UK Atomic Energy Authority, Culham Campus, Abingdon, Oxfordshire OX14 3DB, United Kingdom}
    \affiliation{Department of Engineering Science, University of Oxford, Oxford, OX1 3PH, UK}

\author{Samanyu Tirumala}
	%\email{samanyu.tirumala@ukaea.uk}
	\affiliation{UK Atomic Energy Authority, Culham Campus, Abingdon, Oxfordshire OX14 3DB, United Kingdom}
    \affiliation{Department of Engineering Science, University of Oxford, Oxford, OX1 3PH, UK}

\author{Prashanth Srinivasan}
	% \email{prashanth.srinivasan@ukaea.uk}
	\affiliation{UK Atomic Energy Authority, Culham Campus, Abingdon, Oxfordshire OX14 3DB, United Kingdom}

\author{Ville Jantunen}
	% \email{ville.jantunen@ukaea.uk}
	\affiliation{UK Atomic Energy Authority, Culham Campus, Abingdon, Oxfordshire OX14 3DB, United Kingdom}

\author{Max Boleininger}
	%\email{max.boleininger@ukaea.uk}
	\affiliation{UK Atomic Energy Authority, Culham Campus, Abingdon, Oxfordshire OX14 3DB, United Kingdom}


\begin{abstract}
We use molecular dynamics simulations to directly compute the effective diffusivity of hydrogen gas atoms in homogeneous distributions of monovacancies in tungsten and vanadium, and voids in tungsten. 
Rather than fitting the results to an Arrhenius law, we compare to an analytic approximation for the effective diffusivity recently derived for multi-occupancy traps [Kaur \textit{et al} (2025), Phys. Rev. Mater. \textbf{9}:125404].
We find good agreement between full atomistic simulation and our theory, validating the analytic model for diffusivity for materials containing nanoscale defects characteristic of radiation damage.
There are no parameters fitted, only physically motivated quantities that can be computed with static density functional or atomistic potential calculations. In this study we prove rapid convergence of hydrogen trap occupation to the steady state using lattice kinetic Monte Carlo, the spontaneous emergence of voids in tungsten using atomistic simulation with empirical potentials, and molecular hydrogen formation in voids using molecular dynamics. We conclude with a prediction for diffusion and retention of hydrogen in voids in tungsten starting from first principles.
This work shows that not only is the analytic form for diffusivity and retention in multi-occupancy traps a practical scheme for making predictive simulations of hydrogen isotope diffusion and retention in irradiated microstructures, derived and parameterized from first principles, it is superior to existing single-occupancy trap formalisms.
\end{abstract}

\maketitle

%%%%%%%%%%%%%%%%%%%%%%%%%%%%%%%%%%%%%%%%%%%%%%%%%%%%%%%%%%%%%%%%%
%%%%%%%%%%%%%%%%%%%%%%%%%%%%%%%%%%%%%%%%%%%%%%%%%%%%%%%%%%%%%%%%%
%%%%%%%%%%%%%%%%%%%%%%%%%%%%%%%%%%%%%%%%%%%%%%%%%%%%%%%%%%%%%%%%%

\section{Introduction} \label{sec:introduction}

\noindent
Fusion power promises an abundant source of low-carbon energy, with D-T tokamak fusion power plant designs such as ITER currently the most advanced. While deuterium is plentiful, tritium is a scarce resource, with no more than three kilograms produced annually in heavy water reactors~\cite{tritium-production,tritium-supply-demand}. Understanding the tritium fuel cycle, and minimising losses in the form of tritium retention in irradiated structural materials is therefore critical.
\\

\noindent
In this work, we consider the diffusion-retention properties of hydrogen, in two different materials of interest for fusion power.
Tungsten is the leading candidate material for
plasma facing components due to its combination of high melting temperature and thermal conductivity, and its low thermal expansion~\cite{Federici_FED_2019,You2016} and physical sputtering yield ~\cite{Federici2001}. It also has a radioactive decay sequence under neutron-capture induced transmutations~\cite{GilbertNF2011}, compatible with low-activation requirements. We also consider vanadium, touted as a candidate structural material in lithium-based breeder blanket design. This is owing to its high radiation resistance~\cite{vanadium_rad}, low neutron activation, high-temperature strength~\cite{vanadium_hightemp} and most importantly compatibility with liquid lithium~\cite{vanadium_compat}. 
\\

% \noindent
% \emph{Why vanadium is an important material to consider}
% \\

\noindent
It is reasonably straightforward, at least conceptually, to fit the outgassing measured by Thermal Desorption Spectroscopy (TDS) experiment using the well-known McNabb-Foster~\cite{McNabbFoster_AIME1963} diffusion-retention model, changing parameters representing single-occupancy trap densities and binding energies to reproduce the measured outgassing curves~\cite{Emilio,festim}. If care is taken not to overfit the results~\cite{Akaike_IEEETransAutoControl1974}, such modelling has the benefit of demonstrating compatibility with characteristic binding energies, which can be matched to known defect types.
But using the McNabb-Foster model requires care when the occupancy of the traps is not small.
We can reproduce the maximum retention expected of a multi-occupancy trap with a single-occupancy trap by adjusting the trap density, or by using multiple single-occupancy traps, but we cannot simultaneously match the diffusivity changes~\cite{Kaur_PRM2025}.
This is not necessarily a problem where the gas loading is from an electrolyte into a bcc metal~\cite{gas_electrochemical_permeation_comparison, multi_step_absorption_desorption}, and so the concentration of hydrogen atoms is likely to be below that of impurity traps.
But in plasma loading scenarios, such as might be expected in the divertor~\cite{Delaporte-Mathurin_2024}, this low occupancy assumption is certainly not valid. 
%The effective diffusivity of a low gas concentration is much lower than that at high gas concentration, so under high flux loading conditions the gas concentration will gradually move forward as a sharp front, rather than a slowly-varying interface; the rate of propagation of the front will therefore be determined by the effective diffusivity change as the gas concentration reaches the trap density.
The effective diffusivity is concentration dependent--a low gas concentration may diffuse much more slowly than a high gas concentration, so under high flux loading conditions the gas concentration will gradually move forward as a sharp front, rather than through Fickian diffusion. The rate of propagation of the front will therefore be determined by the effective diffusivity change as the gas concentration reaches the trap density.
In lab-based experimental work, this slow movement of the concentration front is rarely explored explicitly~\cite{VUORIHEIMO2026}, rather the loading is often continued until saturation ~\cite{Schwarz-Selinger_MatResExp2023}.
In modelling the divertor of a fusion power plant, where full-scale experimental results are not yet available, the propagation of this transient, how it may interact with the temperature gradient~\cite{Dasgupta_2023}, and the evolution of defects~\cite{Kato_2015,Lindblad_JNM2025}, must be treated explicitly, and ideally with a parameter-free first-principles model rather than one fitted to the narrow parameter-space window of a limited number of TDS experiments.
\\

\noindent
In this work, we seek to verify our implementation of recently-derived partial differential equations for effective diffusivity of hydrogen gas with multi-occupancy traps typical of irradiation damage, by direct comparison to lattice kinetic Monte Carlo simulations, and to validate the approximations used by direct comparison to molecular dynamics simulations.
This approach is not quite the gold standard of comparison to experiment, but does have the advantage of being able to test equations against a known defect distribution without unknown impurities.
This work differs from previous studies incorporating hydrogen multi-occupancy traps in that we do not fit diffusion rates to MD simulations containing multi-occupancy traps, as in refs ~\cite{Piaggi_JNM2015,Liu_JNM2014,Wang_IJHE2020}, or fit diffusion-retention parameters to experiment ~\cite{zibrov2026damagedosedependencedeuterium}, but instead derive the diffusivity and retention explicitly from first principles calculations.
\\

\noindent
Previous work using massively-overlapping cascade simulations~\cite{Gra15,Gra20,Gra21} has shown that the characteristic microstructure produced by irradiation at low temperature is a near-homogeneous distribution of monovacancies. 
The validity of these microstructures has been shown by comparison to x-ray strain measurements~\cite{Mason_PRL2020}, thermal conductivity~\cite{Mason_PRM2021b}, transmission electron microscopy~\cite{Mason_Acta2023}, and deuterium retention in the saturation limit~\cite{Mason_PRM2021}.
Importantly, dislocation loops have been shown in these studies to offer a limited number trapping sites for hydrogen isotopes~\cite{Boleininger_SciRep2023}, and the trapping energy of loops is not as high as for monovacancies~\cite{DeBacker_PhysScr2017,DeBacker_NucFus2018}.
Therefore an idealised model of irradiated microstructure at low temperature, from the perspective of hydrogen retention, has monovacancies only.
At high temperature, some vacancies will coalesce into voids~\cite{Wang_Acta2023}, while others annihilate on interstitial dislocation loops~\cite{Was_fundamentalsBook}.
A simple idealised model of irradiated microstructre at intermediate temperatures, if considering hydrogen retention and diffusion properties only, has small vacancy clusters and voids only.
Both monovacancies and voids are known to be multi-occupancy traps, with maximum retention expected to scale with the surface area~\cite{Hou_NatMat2019}.
We develop the theory of measured diffusivity and the finite-element simulation of a single isotope in multi-occupancy point-like traps in section \ref{sec:theory}, and validate the analytic form of our equations using kinetic Monte Carlo (kMC) simulations in section \ref{sec:KMC_H_simulation}.
In section \ref{sec:MD_simulation} we describe molecular dynamics (MD) simulations to find the diffusivity of gas atoms in monovacancy distributions, including the full effects of gas-gas interactions, elastic strain, and lattice vibrations, to validate the approximations used in the analytic theory. 
\\

\noindent
The accurate sizing of small nano-voids is difficult by experiment~\cite{Jenkins2001}, so in section \ref{sec:KMC_simulation} we describe simple kinetic Monte Carlo simulations of vacancies migrating on a lattice, to establish a plausible range of vacancy cluster sizes. This study confirms that carbon or other impurities are not required for void growth in tungsten (otherwise we should also need to consider them in our diffusion-retention study).
While it is legitimate to argue the potential energy landscape might be improved, or the defects do not perfectly represent a real material, nevertheless the diffusion computed using Hamiltonian dynamics is a ground truth value for that system. 
Finally in section \ref{sec:voids}, we develop the multi-occupancy model to handle vacancy clusters and voids at finite temperature, and show the agreement between predictive theory and the ground-truth MD simulation.
\\

\noindent
The principle result of this paper is a set of analytic expressions for diffusion and retention of gases in monovacancies and voids, verified by comparison to atomistic kinetic Monte Carlo simulations and validated with full scale Molecular Dynamics simulations.
The good agreement we find between MD and analytic theory cannot easily be reproduced with single occupancy trap modelling, and recommends the use of multi-occupancy traps for future predictive work.
\\

\section{Diffusion-retention equations in multi-occupancy traps}
\label{sec:theory}

\noindent
We start with a simple model of hydrogen isotopes moving through a crystal with defect trap sites. The population of gas atoms is divided conceptually into mobile gas atoms moving from one interstitial site to the next, and trapped gas atoms bound--and so localised--to a point-like defect. A mobile atom can become trapped if it moves within range of a trap, and conversely a trapped atom may be thermally activated from its defect and rejoin the mobile population. Two mobile atoms may not occupy the same interstitial site. A trapped atom may not immediately enter a different trap without passing through a mobile site (note that this assumption makes the following derivation valid for point-like rather than extended defects). It is assumed that the interaction between mobile gas atoms is negligible \cite{Becquart_JNM2009b}. These assumptions are explicitly tested in this paper. 
\\

\noindent
Hydrogenic gas atoms in interstitial lattice sites can diffuse quickly over small activation barriers. 
The diffusion constant measured depends on context.
The flux of mobile gas atoms unimpeded by traps is given by Fick's first law: if $x$ is the mobile gas density, expressed as an atomic fraction, and $\Omega$ the volume per host atom, then
    \begin{equation}
        \label{eqn:Fick1}
        \vec{J}_{x}^{\,\rm{ideal}} = - \frac{D_0}{\Omega} \nabla x.
    \end{equation}
By Einstein's relation~\cite{Einstein1905}, the diffusivity of the mobile gas can be measured from the gradient of the mean square displacement of the mobile atoms with time. In three dimensions, if all atoms are equally mobile, we find the ideal diffusion constant for the interstitial gas:
    \begin{equation}
        D_0 = \lim_{t\rightarrow \infty} \frac{ \langle msd(t) \rangle_x }{6 t} .
    \end{equation}
But if we compute the mean square displacement in a simulation where some atoms are trapped, then we actually measure an effective diffusion constant which can be dramatically lower than the ideal diffusion constant, because only a fraction of the atoms are moving at any time. 
The total gas content, trapped and untrapped, assuming trap types with density $\{\rho_j\}$, can be written $c = x + \sum_j \rho_j \langle \theta_j \rangle$ where $\langle \theta_j \rangle$ is the expected number of trapped gas atoms per trap of type $j$.
    \begin{eqnarray}
        \label{eqn:EffDiff_total}
        D_{\rm eff} &\equiv&  \lim_{t\rightarrow \infty} \frac{ \langle msd(t) \rangle_c }{6 t} = \frac{x}{c} D_0. 
    \end{eqnarray}
\\

\noindent
Viewed on a timescale large in comparison to characteristic trapping and detrapping times, the interstitial gas can be assumed to be in equilibrium with the gas in the traps--known as the Oriani approximation~\cite{Oriani_ActaMet1970}. The range of validity of this assumption is explicitly tested below.  
Under these assumptions, we find that the measured effective diffusivity in equation \ref{eqn:EffDiff_total} can be written
    \begin{equation}
        \label{eqn:EffDiff_mobile}
        D_{\rm eff} = \left( 1 + \frac{\sum_j \rho_j \langle \theta_j \rangle}{x} \right)^{-1} D_0.
    \end{equation}
However, if we take the time derivative of the total gas content (assuming trap density evolution is slow by comparison), then we find~\cite{Kaur_PRM2025}
    \begin{eqnarray}
        \frac{{\rm d} c}{{\rm d} t} &=& \frac{{\rm d} x}{{\rm d} t} + \sum_j \rho_j \frac{{\rm d} \langle \theta_j \rangle}{{\rm d} t} 
        = \frac{{\rm d} x}{{\rm d} t} + \sum_j \rho_j \frac{\partial \langle \theta_j \rangle}{\partial x} \frac{{\rm d} x}{{\rm d} t}  \nonumber\\
        &=& \left( 1 + \frac{\sum_j \rho_j \, {\rm var}\left( \theta_j \right)}{x} \right) \frac{{\rm d} x}{{\rm d} t}.
    \end{eqnarray}
Since the total gas content can only change by diffusion of the mobile population, ${\rm d}c/{\rm d}t = D_0 \nabla^2 x$, we find an equation which can be solved in finite element simulations expressed only in terms of the mobile gas content:
    \begin{equation}
        \label{eqn:EffDiff_mobile}
        \left( 1 + \frac{\sum_j \rho_j \, {\rm var}\left( \theta_j \right)}{x} \right) \frac{{\rm d} x}{{\rm d} t} = \nabla \cdot \left( D_0 \nabla x \right).
    \end{equation}
There is no simple corresponding diffusivity for this expression, but we can define a related single voxel effective diffusivity by 
    \begin{equation}
        \tilde{D}_{\rm eff} \equiv \left( 1 + \frac{\sum_j \rho_j \, {\rm var}\left( \theta_j \right)}{x} \right)^{-1} D_0.
    \end{equation}
Physically, equation \ref{eqn:EffDiff_mobile} means that where the variance is high, it is easy to move H atoms into and out of traps, and hence the diffusivity is low. Conversely, when the variance is low, the traps are effectively invisible to passing mobile H atoms, and the diffusivity is high.
\\

\noindent
If we are measuring the total mean square displacement because we cannot easily separate trapped from mobile populations, we are measuring equation~\ref{eqn:EffDiff_total}. 
This is likely to be the case where we want to compare to atomistically resolved models such as lattice kMC or MD, or to experiment.
Importantly, the effective diffusivity computed from the mean square displacement in such an atomistic simulation, equation~\ref{eqn:EffDiff_total}, is \textit{not} the effective diffusivity needed to solve for the diffusion of the mobile gas under the assumption of local equilibrium in a finite element calculation, where we should instead use equation~\ref{eqn:EffDiff_mobile}.
\\

\noindent
As a concrete example of how to compute the effective diffusion constant, consider the case of a single hydrogen isotope in a multi-occupancy trap.
Writing the probability of finding a trap containing $i$ H atoms as $y_i$, with $i \in \{0,1,\ldots N\}$, we can write the rate of change of probability as a sum of rates of trapping and detrapping events, of the form
    \begin{eqnarray}
        \label{eqn:transient_solution}
        \frac{{\rm d} y_0}{{\rm d} t} &=& - x k \, y_0 + p_1 \, y_1 \nonumber\\
        \frac{{\rm d} y_1}{{\rm d} t} &=& x k \, y_0 - x k \, y_1 - p_1 \, y_1 + p_2 \, y_2 \nonumber\\
        &\ldots&
    \end{eqnarray}
where $k = g f$ is the trapping rate for a mobile gas atom next to a trap, expressed in terms of mobile gas hopping frequency $f$ and geometric factor $g$.
The hopping frequency is $f = \nu \exp\left[-\frac{E^\mathrm{m}}{k_B T} \right]$, where $\nu$ is the attempt frequency and $E^\mathrm{m}$ the migration barrier. 
$p_i = i \, g'_i  f \exp \left[-\frac{E^\mathrm{b}_i}{k_BT} \right]$ is the detrapping rate from a trap with occupancy $i$, where $E^\mathrm{b}_i$ is the $i^\mathrm{th}$ incremental binding energy of a gas atom to the trap, and $g'_i$ is a trap-specific geometric/entropic factor.

The binding and migration energies and attempt frequencies of gas atoms at each occupancy level can be computed with density functional theory or a machine-learned or empirical potential.
A compact notation for the linear set of rate equations in eqn \ref{eqn:transient_solution} is ${\rm d} {\bf y}/{\rm d}t = - {\bf G} {\bf y}$.
In our single isotope example, we can write ${\bf G}$ as a function of mobile gas fraction $x$ and temperature $T$ as
    \begin{equation}
    \label{eqn:tridiagG}
    {\bf G}[x, T] = \left( \begin{array}{ccccc}
                    x k     &   -p_1        &               &             &    0        \\
                    -x k    &   x k + p_1   &   -p_2      &             &             \\
                            &   -x k        &   x k+ p_2   &   -p_3    &             \\
                            &               &   -x k        &    \ddots   & - p_N      \\
                        0   &               &               &    -x k     &  p_N 
                \end{array}
                \right).
    \end{equation}      
\\

\noindent
The multi-occupancy trap equations in their full transient form, equation \ref{eqn:transient_solution}, can be solved numerically with e.g. \texttt{FESTIM2} \cite{festim2}. We note that because matrix ${\bf G}$ has large eigenvalues~\cite{Kaur_PRM2025}, the timestep required for evolution can be quite small, but it is beyond the scope of this work to make a full assessment of the relative computational efficiency of working with transient versus steady state solutions in finite element modelling.
\\

\noindent
The dynamic steady state is defined by the solution to ${\bf G} {\bf y^{\rm eq}} = 0$.
The steady state solution is given by
\begin{equation}
    \label{eqn:steady_state_solution}
    y^\mathrm{eq}_i = \frac{\prod_{j=1}^i x k/p_j}{1 + \sum_{j'=1}^N \prod_{j=1}^{j'} x k/p_{j}},
\end{equation}
A detailed derivation of this steady state, its application, and extension to multiple isotopes can be found in ref~\cite{Kaur_PRM2025}.
The total expected retained gas per trap, also needed for the measured effective diffusivity in atomistic simulation via equation~\ref{eqn:EffDiff_total}, is $\langle \theta \rangle = \sum_i i \, y^{\rm eq}_i$.
The variance, needed for the effective diffusivity for FEM simulation via equation~\ref{eqn:EffDiff_mobile}, is ${\rm var}( \theta ) = \sum_i i^2 \, y^{\rm eq}_i - (\sum_i i \, y^{\rm eq}_i)^2$.
\\

\subsection{Verification with lattice kinetic Monte Carlo}
\label{sec:KMC_H_simulation}

\noindent
In this section we verify the correct implementation of the diffusion in the steady state form, equation \ref{eqn:EffDiff_total}, by direct comparison to lattice kinetic Monte Carlo. Consider a base-centred cubic (bcc) single crystal with periodic boundaries, where the mobile hydrogen atoms move on tetrahedral interstitial sites. We include a vacancy as a prototypical multi-occupancy trap characteristic of low-temperature irradiation damage. Hydrogen atoms are bound to the six octahedral $[ \minifrac{1}{2} \, 0 \, 0 ]$ sites surrounding the monovacancy~\cite{Heinola_PRB2010b}. We assert that the 24 nearest-neighbour tetrahedral $[\minifrac{1}{2} \, \minifrac{1}{4} \, 0]$ sites surrounding the monovacancy are  unstable, so that a hydrogen atom in one of these sites will immediately fall into an octahedral bound site, if one is available. The 24 next-nearest neighbours $[\minifrac{1}{2} \, \minifrac{3}{4} \, 0]$ are regular tetrahedral sites. Note there are four paths from an octahedral site through unstable nearest-neighbour sites to each stable tetrahedral next-nearest-neighbour site--this detrap rate prefactor is included in the lattice kMC. The trapping and detrapping rates for the lattice kMC simulation are now defined by this geometry and the binding/migration energies. 
Here we use the energies for hydrogen in tungsten, excluding zero-point energies using the empirical EAM potential MNL2023~\cite{Mason_JPCM2023}. The trap binding energies are given in ref~\cite{Mason_JPCM2023} and repeated here in table \ref{tab:binding_energy} for convenience together with DFT values for comparison. We may conclude from this table that the empirical potential values are not an unphysical choice, which means we can also compare to full molecular dynamics simulations below.
\\

\begin{table}[htb!]
    \centering
    \begin{tabular}{c|l|l|l}
        Value           &   EAM &       DFT     &   units\\
        \hline
         $\Delta E^\mathrm{m} $    &   0.216       &    0.21 $^a$    & eV\\
         $\Delta E_1^\mathrm{b}$   &   1.233       &    1.28 $^b$    & eV\\
         $\Delta E_2^\mathrm{b}$   &   1.349       &    1.25     & eV\\
         $\Delta E_3^\mathrm{b}$   &   0.975       &    1.11     & eV\\
         $\Delta E_4^\mathrm{b}$   &   1.050       &    1.00     & eV\\
         $\Delta E_5^\mathrm{b}$   &   0.750       &    0.91     & eV\\
         $\Delta E_6^\mathrm{b}$   &   0.486       &    0.32     & eV\\
         $E^\mathrm{f}( H )$       &   0.798       &    0.685 $^a$   & eV\\
         $\nu$            &   38          &   26.2 $^c$ & THz
    \end{tabular}
    \caption{Low temperature migration and binding energies for hydrogen in tungsten, excluding zero point corrections, parameterized using the MNL2023 EAM potential~\cite{Mason_JPCM2023}. The binding energy is defined in terms of vacancy+H cluster formation energy and the heat of solution as $\Delta E_i^\mathrm{b} = E^\mathrm{f}( \mathrm{v H}_{i-1} ) + E^\mathrm{f}( \mathrm{H} ) - E^\mathrm{f}( \mathrm{v H}_i )$. 
    DFT values from $^a$ Ref~\cite{Heinola_PRB2010}, $^b$ Ref~\cite{Heinola_PRB2010b}, $^c$ Ref~\cite{Heinola_JAP2010} for comparison.
    }
    \label{tab:binding_energy}
\end{table}

\noindent
We use a system size $200\times 200\times 200$ conventional bcc unit cells (16M tungsten atoms), with 16000 hydrogen atoms (0.1 at \%)  and a varying count of vacancies. We use the n-fold way~\cite{Bortz_JCP1975} to select a single H atom to move each time step, and increment the system clock at each step by a time $\delta t$ equal to the inverse of the sum of rates. 
\\

\begin{figure}
    \centering
    \includegraphics[width=0.9\linewidth]{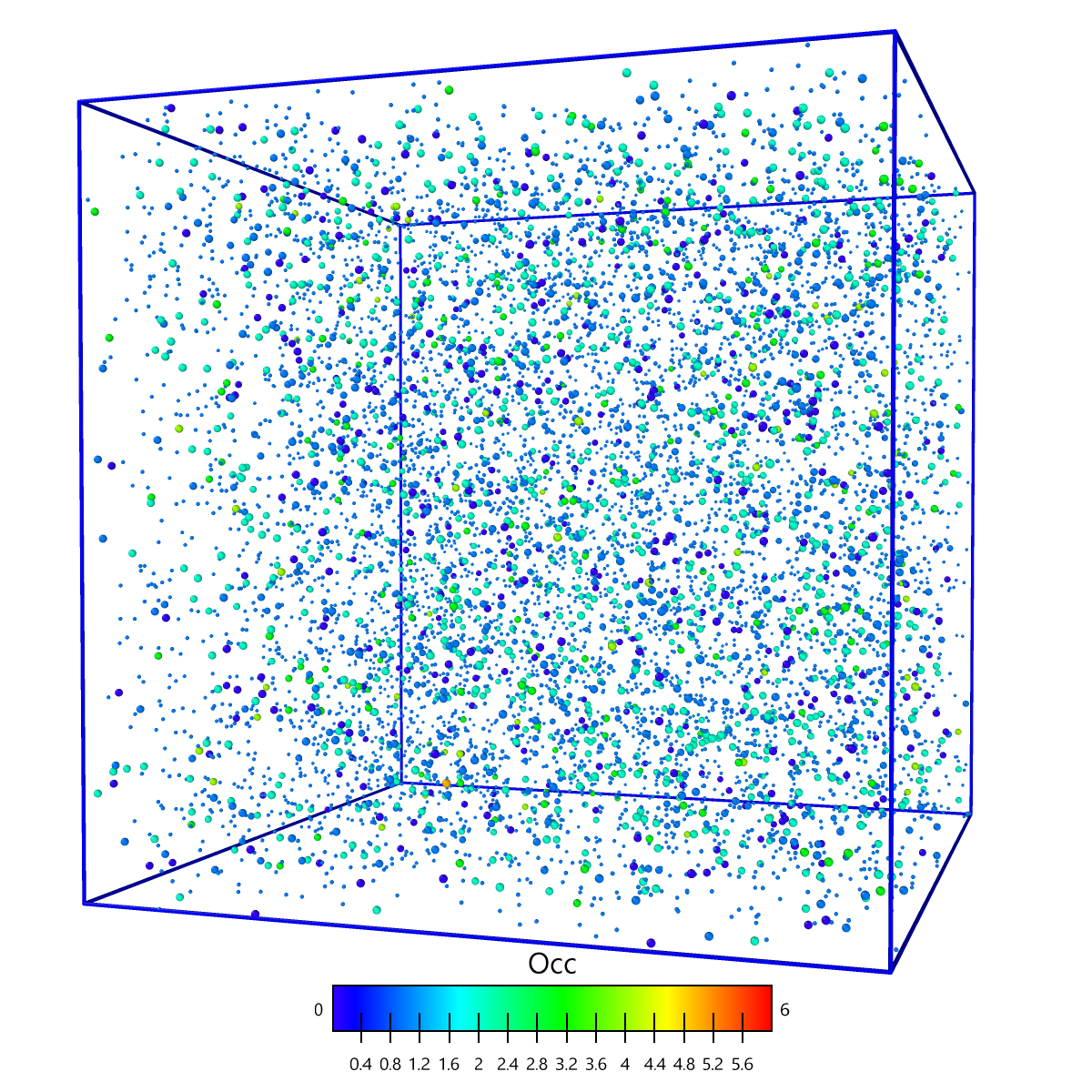}
    \caption{Snapshot from lattice kMC simulation with 16M tungsten atoms (not shown), 16k hydrogen atoms, and 4.8k vacancies after 10000 hops per H atom at 1500K. Mobile H is shown as small dots. Vacancy traps shown as spheres, coloured by occupation. Visualisation with Ovito~\cite{Stukowski_MSMSE2009}.
    }
    \label{fig:kmc_H_in_vac}
\end{figure}

\begin{figure*}
    \centering
    \includegraphics[width=0.95\linewidth]{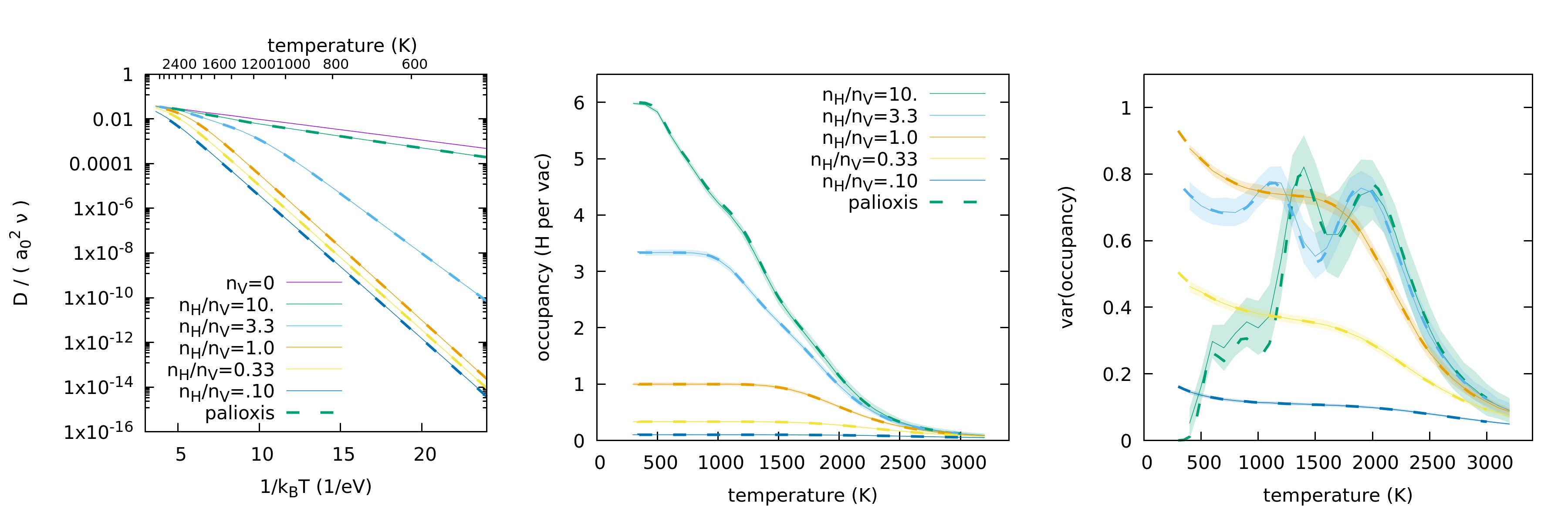}
    \caption{Diffusion constant, average occupancy per vacancy trap, and variance of occupancy per trap. Lattice kMC results shown as solid lines plus shaded regions indicating the standard deviation of results. Steady state analytic results generated with Palioxis \cite{Kaur_PRM2025,Palioxis} are shown as dashed lines.}
    \label{fig:compare_palioxis_kmc}
\end{figure*}

\noindent
To compare this simulation directly with our steady state solution, note that the ideal diffusion constant for hopping between tetrahedral sites is $D_0 = 4 f a^2 /6$, with $a$ being the distance between tetrahedral sites, $a=a_0/\sqrt{8}$. The trapping rate geometric factor is $g=4$, as there are 24 stable tetrahedral next-nearest neighbour sites which can trap into 6 octahedral sites.
We set the detrapping rate geometric factor to $g'_i = 4g$. We explain this choice below when comparing different physical approximations, for now it is sufficient to recognise we simply want to compare models, and the specifics of binding energies and rate constants made are irrelevant details for this part of the study. Changing to $g'_i = g$ has no effect on the conclusion.
Measuring the diffusion constant of the hydrogen atoms in lattice kMC directly from their mean square displacement as $D_{\rm eff}= \lim_{t\rightarrow\infty} \langle msd(t)\rangle_c ⁄6t$, where the mean square displacement at time $t$ has the drift removed~\cite{Derlet_PRB2011}, we find in convergence studies that the steady state is reached at order 1000 hops per hydrogen atom--this convergence is demonstrated in appendix~\ref{sec:convergence_soas_H_in_v}. We compute properties of the kMC system- the diffusion constant $D_{\rm eff}$, mean occupancy $\langle \theta\rangle$, and variance of occupancy $\rm{var}(\theta)$.
Starting with hydrogen atoms in randomly-assigned tetrahedral sites at 3200K, we burn in for 10000 hops per H atom, then sample for 2000 hops per H atom. We then reduce the temperature by 100K, and repeat. A snapshot from a simulation at 1500K is given in figure \ref{fig:kmc_H_in_vac}. Results are shown in figure \ref{fig:compare_palioxis_kmc}, with shaded regions indicating the standard deviation of results found over 12 independent simulation runs. 
The complex behaviour of the variance of the occupancy is due to the non-monotonic incremental binding energy in this choice of potentials.
\\

\noindent
We see that analytic steady-state results, computed using the \texttt{PALIOXIS} code~\cite{Palioxis}, are a very good match to the fully time- and spatially- dependent kMC result, verifying our analytic model indeed recovers steady-state properties.
We note in passing that generating results with the kMC code took a few hours, whereas the results using the steady state computed using equation \ref{eqn:steady_state_solution} took milliseconds.
\\

%%%%%%%%%%%%%%%%%%%%%%%%%%%%%%%%%%%%%%%%%%%%%%%%%%%%%%%%%%%%

\section{MD simulation}
\label{sec:MD_simulation}

\noindent
In this section we make a validation for the multi-occupancy steady-state computed using equation \ref{eqn:steady_state_solution}, with MD. MD represents a step change in the fidelity of the physics represented. Hydrogen atoms can now interact through elastic fields, thermal expansion is present, correlation effects will be present, and simple assumptions of low-temperature binding energies are replaced by the full dynamics including phonon and anharmonic contributions to free energy.
This section therefore tests the validity of the approximations made.
Molecular dynamics simulations were run using {\texttt LAMMPS}~\cite{LAMMPS} to find the diffusivity of hydrogen atoms.
%The simulations were performed in the NPT ensemble, with periods of thermalization followed by periods of sampling the mean square displacement as a function of time. 
We fit the diffusivity with drift removed as above using $D_{\rm eff}=\lim_{t\rightarrow\infty} \langle msd(t)\rangle_c ⁄6t$.
\\

\begin{figure}
    \centering
    \includegraphics[width=0.7\linewidth]{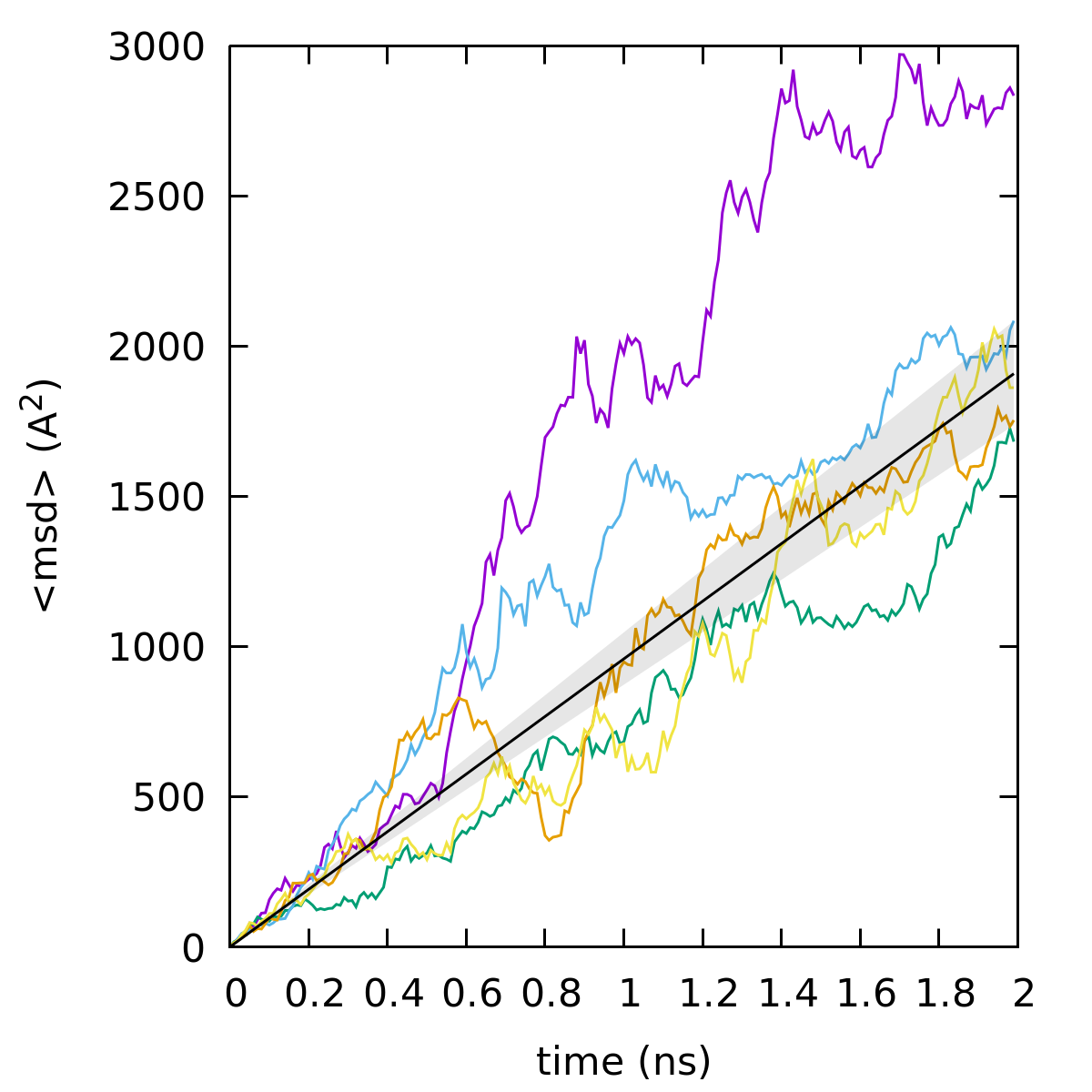}
    \caption{Illustrative calculation of diffusion constant from mean-square displacement over 5 trials computed with MD. Traces taken at 1200K, with 0.03 at \% total H concentration and 0.03 at \% vacancy concentration. The solid black line and shaded region are the best linear fit and 68\% confidence interval.}
    \label{fig:msd}
\end{figure}

\noindent
The simulations performed for hydrogen in tungsten used a simulation cell size of $48\times 48 \times 48$ conventional bcc unit cells--considerably smaller than the lattice kMC simulations above, with varying hydrogen and vacancy concentrations, using the same MNL2023 empirical potential as before. 
Starting at a temperature of 2200K then slowly cooling in 200K steps, we used 5 cycles of 2\,ns thermalization followed by 2\,ns sampling, see figure \ref{fig:msd}.
Simulations were performed in the NPT ensemble. MD simulation took three orders of magnitude longer than the kMC simulations despite this much smaller system size and shorter sampling time.
\\

\noindent
To make an explicit comparison between the MD simulation and the analytic result, we must consider possible physical effects in the choice of the detrapping geometric/entropic factor $g'_i$.
Consider the choice $g'_i = \gamma g/i$, where $\gamma$ is the number of interstitial sites per host atom (eg in bcc the H atoms are on $\gamma=6$ tetrahedral sites per host atom). Then, given the definition of the incremental binding energy,
    \begin{equation}
        E^{\rm b}_i = (E^{\rm f}_{i-1} + E^{\rm f}_{\rm L}) - E^{\rm f}_i,
    \end{equation}
where $E^{\rm f}_i$ is the formation energy of trap plus $i$ H atoms, and $E^f_{\rm L}$ is the formation energy of the H atom in a interstitial lattice site. Substituting into equation \ref{eqn:steady_state_solution}, we find 
    \begin{equation}
        y^{\rm eq}_i \sim \exp\left[ - \frac{E^{\rm f}_i - i \mu_{\rm H}}{k_B T} \right],
    \end{equation}
where $\mu_{\rm H} = E^{\rm f}_{\rm L} + k_B T \ln{(x/\gamma)}$ is the chemical potential appropriate for a mobile fraction $x$. We recognise then that this choice of $g'_i$ gives us the \textit{thermodynamic} steady state.
\\

\noindent
A second choice for $g'_i$ might include entropic effects from phonon or configurational entropy, by multiplying $g'_i$ by $\exp[ - (F^{\rm b}_i - E^{\rm b}_i)/k_B T ]$, where $F^{\rm b}_i$ is an incremental binding free energy, defined equivalently to the incremental binding internal energy above. Phonon effects in the classical approximation can be introduced writing $F^{\rm f} = E^{\rm f} + 3 k_B T \log ( \hbar \omega_i / k_B T )$, where $\omega_i$ is an appropriate average vibrational frequency of a hydrogen atom in a monovacancy trap containing $i$ hydrogen atoms. This gives a phononic contribution to the detrapping rate
    \begin{equation}
        {g'_i}^{\rm therm+ph} = g \, \frac{\gamma}{i} \frac{\omega_{i}^{3i}} {\omega_{i-i}^{3i-3} \omega_{\rm L}^3},
    \end{equation}
where $\omega_{\rm L}$ is the vibrational frequency of an H atom in the interstitial lattice site.
Configurational effects can be added with $F^{\rm f} = E^{\rm f} - k_B T \log ( W_i )$, $W_i$ is a Boltzmann configurational entropy corresponding to the number of degenerate configurations that the hydrogen atoms can adopt, and is appropriate if the swapping between these degenerate configurations is fast compared to the detrapping.
For a monovacancy in a bcc metal, DFT results suggest the number of degenerate states as a function of occupation are $W_i = \{1,6,3,12,12,3,1\}$ for $i \in \{0,\ldots 6\}$.
    \begin{equation}
        {g'_i}^{\rm therm+conf} = g \, \frac{\gamma}{i} \frac{W_{i-1}}{W_i}. 
    \end{equation}
\\

\begin{figure*}
    \centering
    \includegraphics[width=0.6\linewidth]{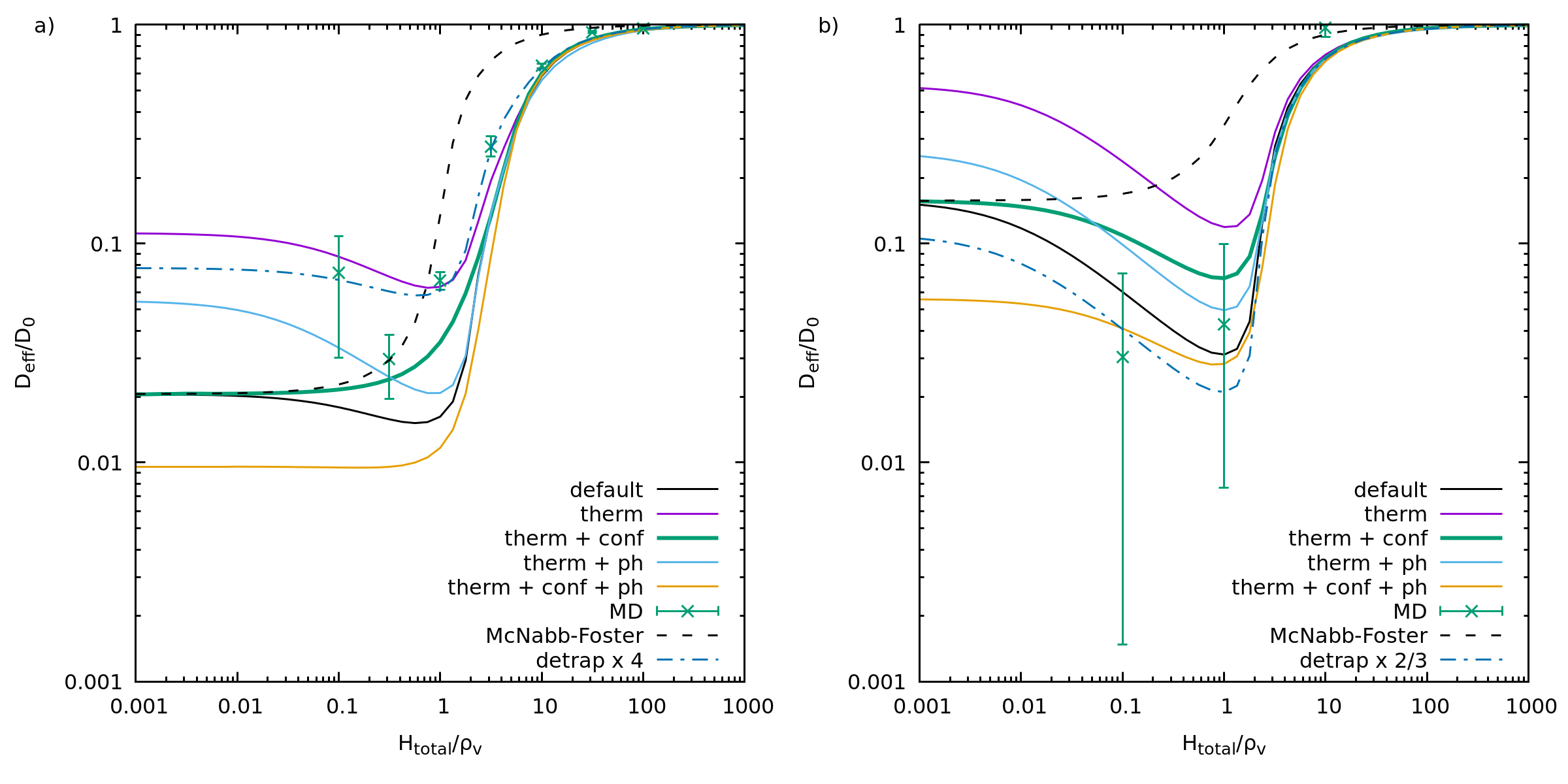}
    \caption{Scaled effective diffusivity of hydrogen in a) tungsten at 1200K and b) vanadium at 500K with monovacancy defects. Points computed with MD, compared to analytic curves using different physical approximations described in the text.}
    \label{fig:compare_physics}
\end{figure*}

\noindent
Figure \ref{fig:compare_physics} shows the effects of introducing these different physics to the detrapping rate, by plotting the effective diffusivity as a function of hydrogen content in the box with multi-occupancy monovacancy traps.
    Figure \ref{fig:compare_physics}a) shows tungsten at 1200K with a vacancy content set to $\rho_{\rm v} = 0.03\%$, and figure \ref{fig:compare_physics}b) shows vanadium at 500K with a vacancy content of $\rho_{\rm v} = 0.1\%$. Changing the functional form of the detrapping rate makes a considerable difference to the predicted diffusivity, but there is no clear `best' form. But this result should not be a surprise, in the MD simulations the hydrogen atoms are not in equivalent sites with equivalent energies and phonon frequencies, and there are elastic effects not included in the simple model.
We can say from figure \ref{fig:compare_physics} that including phonon or configurational entropy effects gives a transferably better answer than not including them. Furthermore, since phonon frequencies are far more difficult to compute or find in tabulated form, we recommend using the configurational entropy only, and in the remainder of this section we use ${g'_i}^{\rm therm+conf}$.
Note that the thermodynamic prefactor $g'_i = \gamma g/i$ was suggested by Hodille et al in \texttt{mHIMS}~\cite{Hodille_PhysScr2016}, the default was suggested for \texttt{TESSIM-x}~\cite{Schmid_JAP2014}, and \texttt{FESTIM-2} leaves the choice to the user.
\\

\noindent
Note that we also plot on figure \ref{fig:compare_physics} an empirically-fitted curve which tunes the detrapping rate to the MD result. For tungsten, we find $g_i \approx 4 g$ works well--this is the number we used in the previous section for the lattice kMC simulation. For vanadium, we find $g_i \approx 2/3 g$. We do not recommend using these empirically-fitted values.
Finally, on figure \ref{fig:compare_physics}, we show the prediction from the McNabb-Foster (single-occupancy trap) model. At very low or very high hydrogen concentrations, we find that McNabb-Foster coincides with the default $g'_i = g$: at low concentrations the traps have $\langle \theta \rangle \ll 1$, so the single-occupancy approximation holds, and at high concentrations, most of the H atoms are untrapped, so the ratio $D_{\rm eff}/D_0$ tends to 1. But in the intermediate range where $c_{\rm H} \sim \rho_{\rm v}$, the McNabb-Foster approximation is a poor fit to the MD results, especially for vanadium.
\\

% For the tungsten simulations, we used the MNL23 W-H potential~\cite{Mason_JPCM2023}, which gives good void properties as well as good binding energies of hydrogen atoms to voids.
% Simulation cells sized $48\times 48 \times 48$ conventional bcc unit cells (221k atoms) were used, with vacancies created by removing atoms at random. 
% The binding energies used to parameterize equation \ref{eqn:tridiagG} are tabulated in ref~\cite{Mason_JPCM2023}.
% \\

\begin{figure*}
    \centering
    \includegraphics[width=0.9\linewidth]{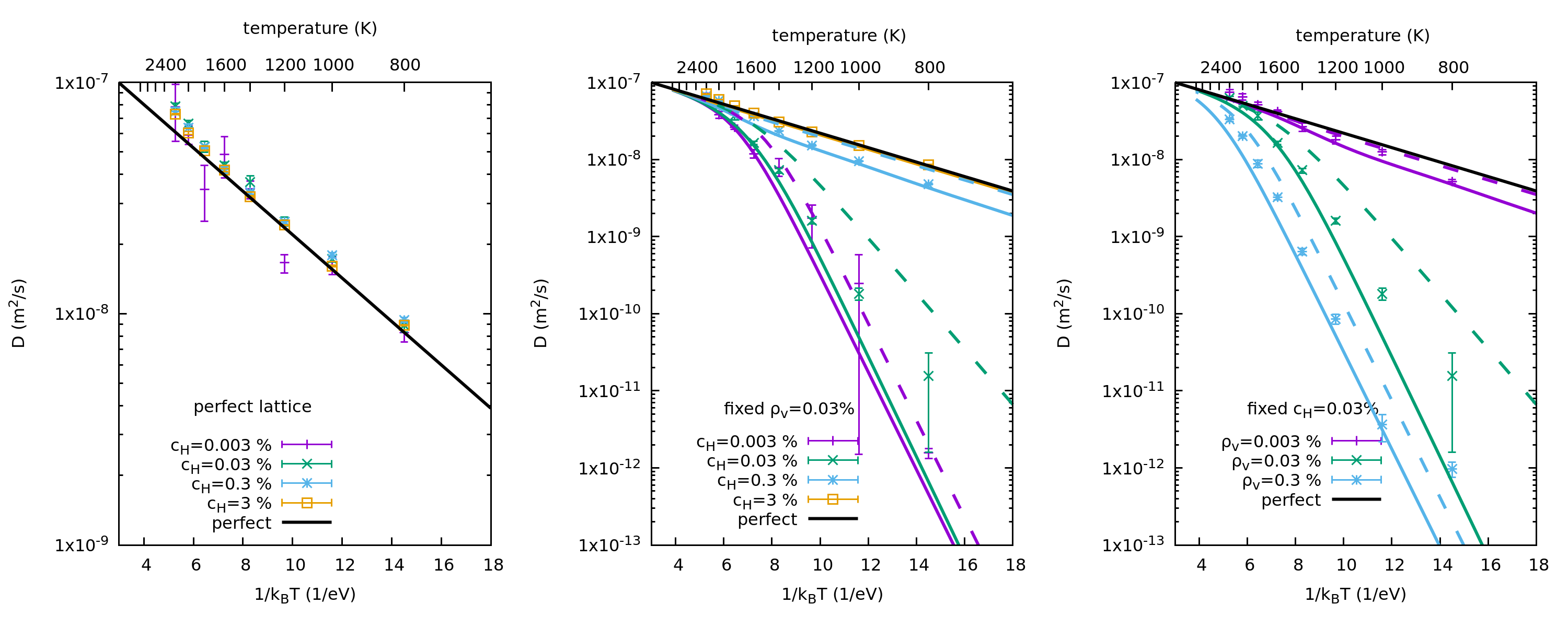}
    \caption{Effective diffusivity of hydrogen in tungsten with monovacancy defects computed with MD, compared to analytic curves. Left: perfect lattice containing no vacancies. Centre: fixed vacancy content 0.03 at\% with varying hydrogen gas concentration. Right: fixed hydrogen concentration with varying vacancy content. Solid lines are analytic multi-occupancy trap approximation, dotted lines show single-occupancy trap approximation.}
    \label{fig:diffusion_MNL-W}
\end{figure*}

\noindent
Results for the diffusivity of hydrogen atoms in tungsten containing monovacancy defects are shown in figure \ref{fig:diffusion_MNL-W}.
The error bars represent 68\% confidence intervals, found by sampling over short sections $\ge 25\%$ of the total diffusion trajectory.
We see no significant difference in diffusivity in the perfect lattice with hydrogen concentration. While tungsten is not hydride forming, and the hydrogen-hydrogen interactions in the lattice are usually considered weak \cite{Becquart_JNM2009b}, nevertheless at these high concentrations some hydrogen clustering has previously been reported \cite{Wang_IJHE2020,Qin_JNM2015}. Here we see no hydrogen concentration effect.
The change in diffusivity for different vacancy and hydrogen concentrations are shown in the central and right plots of figure \ref{fig:diffusion_MNL-W}.  
The diffusion results shown in figure \ref{fig:diffusion_MNL-W} are similar to those found in an MD study of hydrogen diffusion in tungsten by Liu et al~\cite{Liu_JNM2014} using the Li2011 potential~\cite{Li_JNM2011}.
The analytic curves, computed with the same parameterization as used in the lattice kinetic Monte Carlo simulation in section \ref{sec:KMC_H_simulation}, lie very close to the ground-truth MD data points.
\\

% \noindent
% We also consider the Mc-Nabb Foster single-occupancy trap approximation, using one trap with the first binding energy $\Delta E^\mathrm{b}_1$ only. This is shown in figure \ref{fig:diffusion_MNL-W} as a dotted line. As expected, this shows a reasonable fit to the MD data where the trap density exceeds the hydrogen concentration. To show the differences in the single-occupancy vs multi-occupancy approximations more clearly, figure \ref{fig:diffusion_W_at_1200K} shows the variation of diffusion constant as a function of hydrogen content with vacancy count fixed and vice-versa at a single temperature 1200K. We see that the multi-occupancy trap model is indeed a better fit to the MD results than using single occupancy traps.
% \\

\noindent
For the vanadium simulations, we used the atomic cluster expansion (ACE) machine-learning interatomic potential (MLIP)~\cite{srinivasan_vh_prep} which reproduces the interstitial solution energy, the migration energy barrier, and the incremental hydrogen binding energies to \textit{ab initio} accuracy. Simulation cells sized $32\times32\times32$ conventional bcc unit cells having 65,536 atoms were used. For the various simulations, H atoms were randomly inserted, and V atoms were randomly removed. We used 5 cycles of 0.5\,ns thermalization followed by 10\,ns tracking mean square displacement. The binding energies used to parameterize equation \ref{eqn:tridiagG} are tabulated in ref~\cite{srinivasan_vh_prep}.
\\

\begin{figure*}
    \centering
    \includegraphics[width=0.9\linewidth]{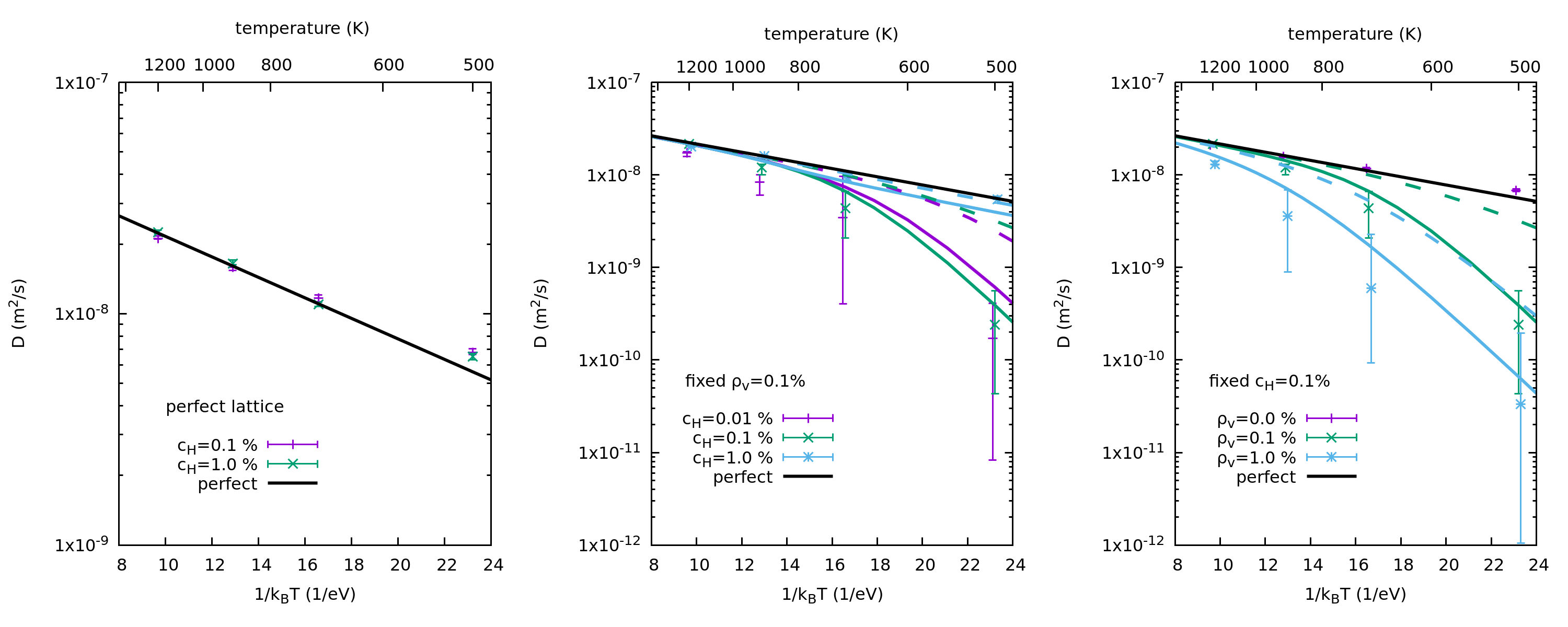}
    \caption{Effective diffusivity of hydrogen in vanadium with monovacancy defects computed with MD, compared to analytic curves. Left: perfect lattice containing no vacancies. Centre: fixed vacancy content 0.1 at\% with varying hydrogen gas concentration. Right: fixed hydrogen concentration with varying vacancy content. Dashed lines: McNabb-Foster approximation.}
    \label{fig:diffusion_ACE-V}
\end{figure*}

\noindent
Results for the diffusivity of hydrogen atoms in vanadium containing different atomic fractions of hydrogen and vacancies are shown in figure \ref{fig:diffusion_ACE-V}.
On the left, we see that for hydrogen diffusing in the perfect lattice, there is no significant difference in the concentration range 0.1--1.0 at \% hydrogen. These concentrations are very high, and so we can conclude that even in this hydride-forming system the effect of hydrogen-hydrogen interactions in the lattice is negligible, and so the system is eligible for modelling using our analytic expressions.
\\

\noindent 
The central plot shows the diffusivity in vanadium with 0.1\% monovacancies for different hydrogen content. Below 0.1\% hydrogen concentration, there is no significant change in the diffusivity even at low temperatures. On the right, we see the effect of vacancy concentration on diffusion of 0.1\% fraction of hydrogen atoms. Higher vacancy content leads to more hydrogen trapping and a subsequent drop in the diffusivity. This effect becomes weaker at higher temperatures, with the diffusivity nearing ideal lattice behaviour. The drop in diffusivity calculated using the analytical model falls very close to the ground-truth MD results, within their uncertainty limit (shown as  68\% standard deviation error bars), thereby validating its use for hydrogen diffusion in vanadium in these concentration limits. The error bars are larger at low temperatures especially when the hydrogen to vacancy ratio is less than 1, owing to the slower rate of detrapping and the subsequent noise in the mean square displacement plots.
We conclude that the approximations used in deriving our analytic expressions for multi-occupancy traps are valid for modelling bcc metals containing monovacancies.
\\
% \noindent
% \emph{something about const cV, const cH calcs}
% \\

\noindent
We also consider the McNabb Foster single-occupancy trap approximation, using one trap with the first binding energy $\Delta E^\mathrm{b}_1$ only. This is shown in figures \ref{fig:diffusion_MNL-W} and \ref{fig:diffusion_ACE-V} with dotted lines. As expected, this shows a reasonable fit to the MD data where the hydrogen concentration is much smaller or much greater than the trap concentration.
Taken for a particular comparable hydrogen and trap concentration, it is easy to fit the McNabb Foster single-occupancy trap approximation to the MD simulation results, by tweaking the detrapping prefactor $g'_1$. But when considered across the two materials considered here and a range of trap and hydrogen concentrations, it is clear that the multi-occupancy trap model is indeed a better fit to the MD results.
 \\

%%%%%%%%%%%%%%%%%%%%%%%%%%%%%%%%%%%%%%%%%%%%%%%%%%%%%%%%%%%%

\section{Retention and diffusion in tungsten nanovoids}

\subsection{KMC void growth simulation}
\label{sec:KMC_simulation}

\noindent
In this section we consider characteristic void cluster sizes suitable for an irradiated material using lattice-based kinetic Monte Carlo.
We assume all atoms are placed into ideal bcc lattice sites, with some lattice sites vacant. The rate at which an atom moves into a vacant site is given by the Kang-Weinberg~\cite{Kang_JCP1989} rule:
    \begin{equation}
        r = \nu \exp\left[ - \frac{\Delta E^\mathrm{m}}{k_B T} \right] \, \exp\left[ - \frac{E_{\rm after} - E_{\rm before}}{2 k_BT} \right],
    \end{equation}
where $\nu$ is the hopping frequency, $\Delta E^\mathrm{m}$ is the migration barrier and $E_{\rm before}$, $E_{\rm after}$ are the energies of the system in the metastable configuration before and after a transition.
The energies are computed using an embedded atom potential; for tungsten we use the MNB17 potential~\cite{Mason_JPCM2017}, which shows good vacancy cluster energies compared to DFT calculations. For vanadium we use the Han23 potential~\cite{Han_JApplPhys2003} which was developed specifically to study radiation damage. Here, we resort to an EAM potential which is roughly 500-1000 times cheaper than the previously used ACE potential, since the kMC simulations are performed on large simulation cells.
The hopping frequency scales the absolute rate linearly, so we choose a conventional approximate value {$\nu=10^{12}$\,Hz} for both materials. Order-1 variations in this value make little difference to the result.
We acknowledge that the Kang-Weinberg rule is highly simplified, and will miss the effect of the anomalous high mobility of trivacancy clusters~\cite{Barouh_PRB2015,Wei_ActaMat2025}. The advantage in using it is that we do not need to perform expensive saddle point energy calculations, and can run optimised vacancy clustering simulations with thousands of vacancy walkers making millions of hops each.
\\

\noindent
The time-evolution is simulated using the rejection-kMC method. 
First, we generate transition rates, $r_j$, for all moves of all atoms neighbouring vacancies. 
This gives a total rate $R_i$ for vacancy $i$ to move as the sum of the rates of neighbouring atoms which can exchange with vacancy $i$. For an isolated monovacancy, there are 8 neighbouring atoms with the same rate; for a vacancy at the surface of a void the atom count is lower, and the rate of each hop is different.
Then, we compute the probability that each vacancy moves in time $\delta t$, given by $p_i = R_i \delta t$.
Moving a vacancy affects not just its own rates, but also others within a range of twice the potential cut-off distance.
If two vacancies were to be selected to move such that each other's rates would be affected, their movement would be correlated.
In this work, we check whether two correlated vacancies are selected in time $\delta t$, and if they are, we reject all moves for this step and reduce the timestep $\delta t$. If no correlated vacancies are selected, we increase the timestep.
When a set of uncorrelated vacancies is selected, one of the neighbouring atoms is selected to actually exchange with the vacancy, with a 
probability proportional to $r_j$.
\\

\begin{figure}
    \centering
    \includegraphics[width=0.45\linewidth]{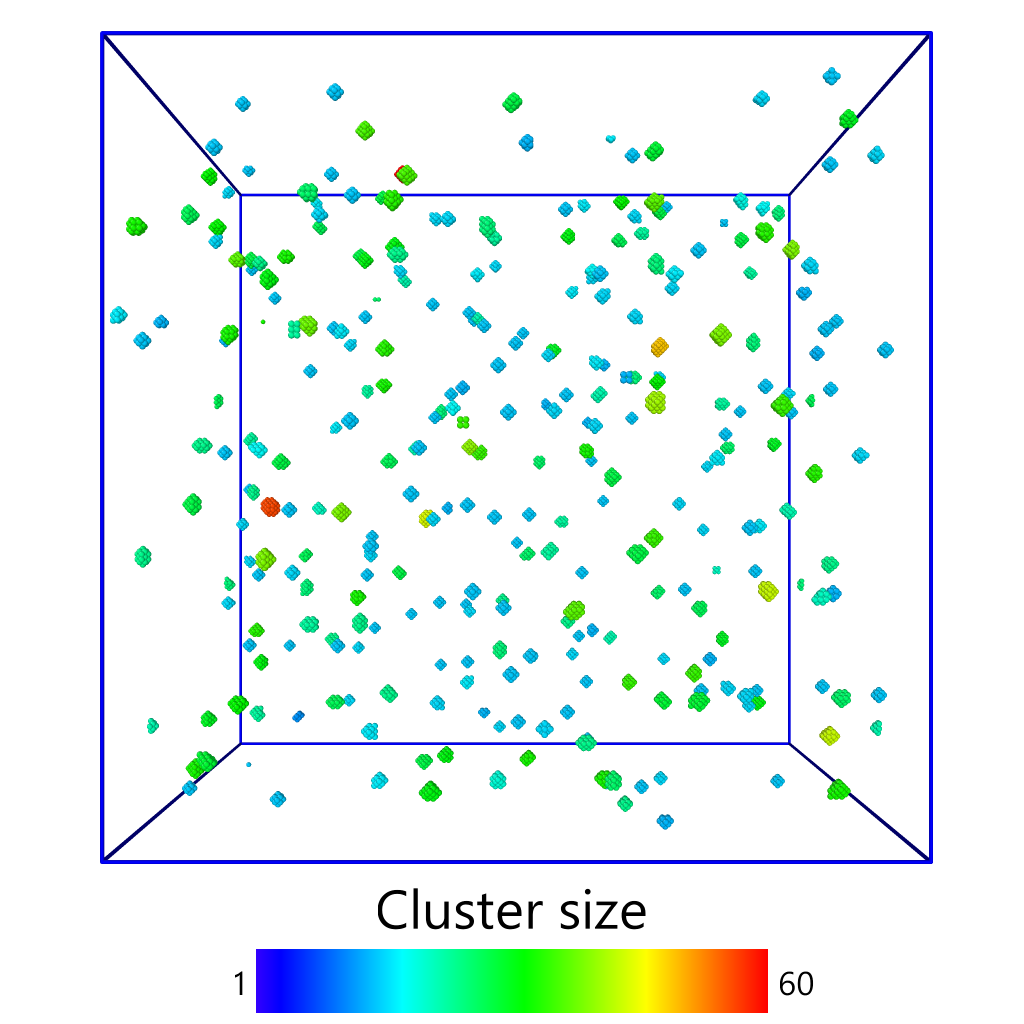}
    \includegraphics[width=0.45\linewidth]{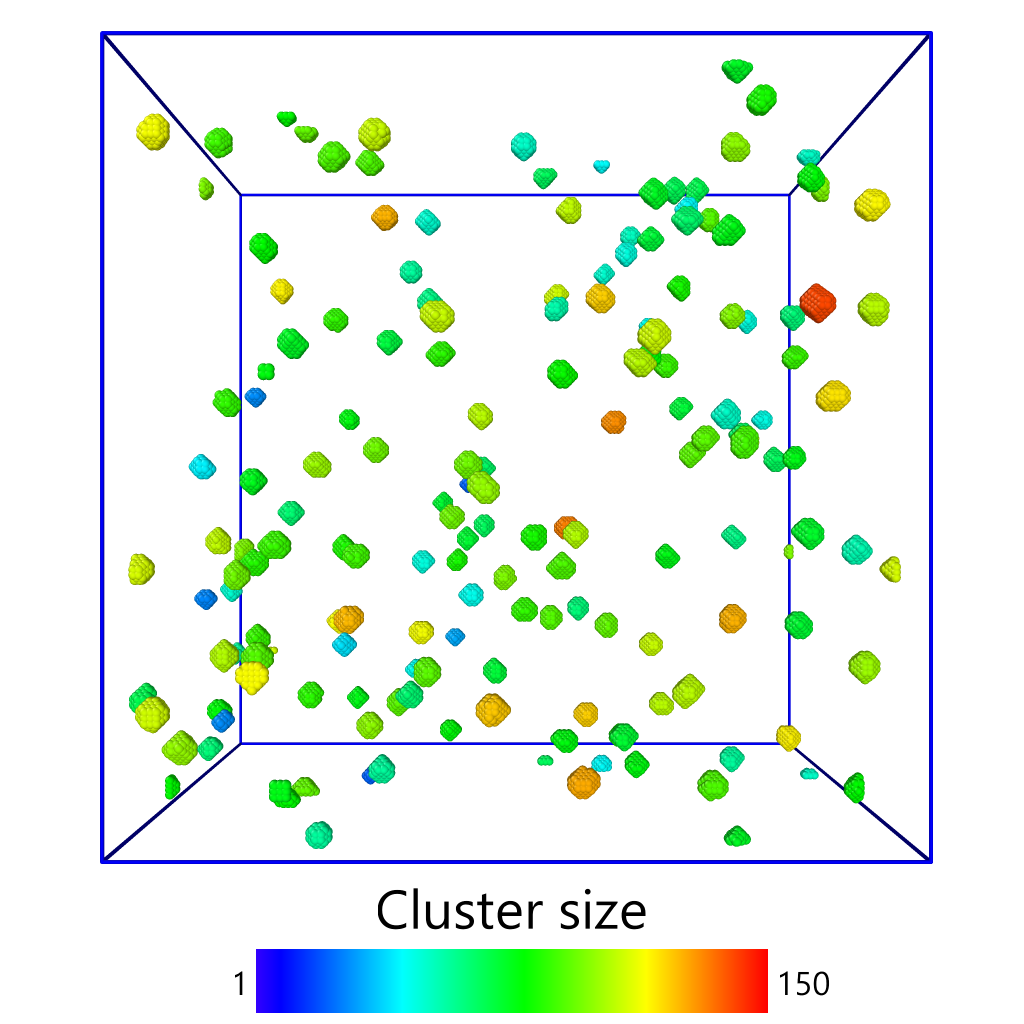}
    \caption{
    Snapshot of kMC vacancy cluster simulations containing 6M atoms (box side order 45nm). Left: vanadium containing 0.1 at\% vacancies at 1200K. Right: tungsten containing 0.2 at\% vacancies at 1200K. Rendered with Ovito~\cite{Stukowski_MSMSE2009}.
    }
    \label{fig:vac_clusters}
\end{figure}

\noindent
We used simulation cells containing $147\times 147 \times 147$ unit cells, with 0.1 at \% to 0.3 at \% vacancies (6k to 19k vacancies). These vacancy concentrations were chosen to represent the range of saturated vacancy concentration in ion irradiated materials \cite{Boleininger_SciRep2023,Gra21}, and so these simulations represent annealing of saturated low-temperature irradiation damage rather than vacancy cluster coalescence during irradiation.
Snapshots from simulations are shown in figure \ref{fig:vac_clusters}. 
\\

\begin{figure}
    \centering
    \includegraphics[width=0.9\linewidth]{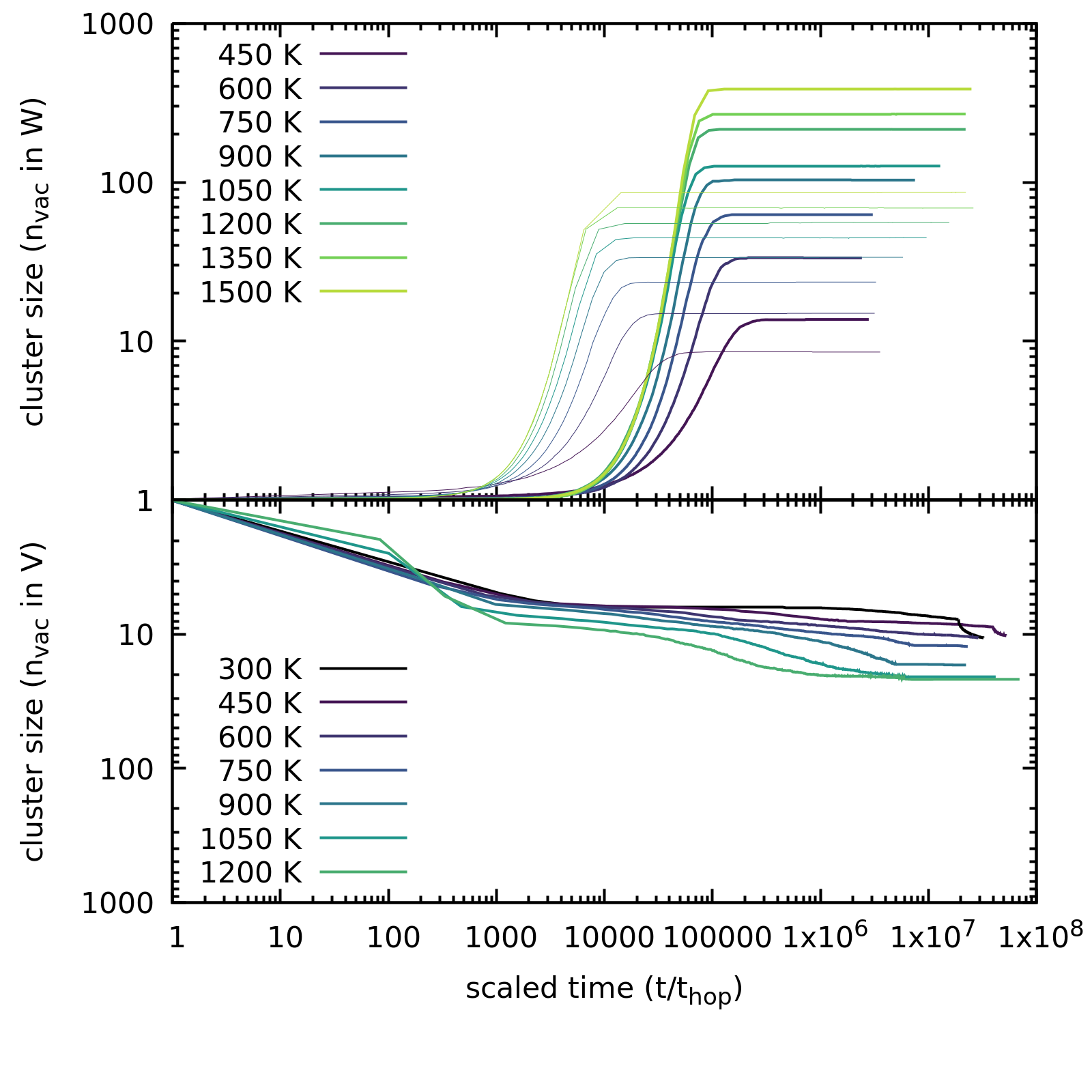}
    \caption{Average void cluster size as a function of number of moves per vacancy for tungsten (above) and vanadium (below). Note the split y-axis. Vacancy content is 0.1 at \% (thick lines) and 0.3 at \% (thin lines).
    }
    \label{fig:vac_clust_W-V}
\end{figure}

\noindent
The vacancy clustering results as a function of time for different annealing temperatures are shown in figure \ref{fig:vac_clust_W-V}.
We note that both materials generate voids via a nucleation and growth mechanism without the need for invoking additional stabilisation through carbon impurities~\cite{Niu_JNM2023,Song_ActaMat2024}, hydrogen atoms~\cite{Lindblad_JNM2025,Qin_JNM2015}, or vibrational free energy corrections~\cite{lapointe2025anomalous}.
\\

\noindent
The main difference between the two interatomic potentials chosen for this study is the binding energy of small vacancy clusters. Tungsten is known to have a near-zero binding energy for the divacancy at low temperature--for the potential used {$E^\mathrm{b} = 0.107$\,eV} compared to DFT {$E^\mathrm{b} \lessapprox 0.05$\,eV}~\cite{Mason_JPCM2017,Muzyk_PRB2011}, while vanadium has a finite binding energy {$E^\mathrm{b} \sim 0.22$\,eV}. This difference leads to earlier production of vacancy clusters in vanadium. In figure \ref{fig:vac_clust_W-V} we see an initial slow growth in vacancy clusters in vanadium up to times order 1000 hops, then a plateauing, then a slow further growth. This second growth phase is due to Brownian motion of small vacancy clusters, which occurs at a slower rate than monovacancy migration. Simulations of iron vacancy clusters using Ackland 97 \cite{Ackland97} (not shown here) shows very similar qualitative behaviour, but slightly larger clusters.
\\

\noindent
In the tungsten case we see little growth before 1000 hops per vacancy, as small clusters are unstable even at low temperatures. Only when a stable nucleus is formed does the cluster grow, and then it does so very rapidly to a first plateau similar to that seen in vanadium. We do not see a second growth phase due to Brownian motion in tungsten, because the clusters are so large that they move very slowly indeed--in simulations we see some rearrangements at the surface but no macroscopic diffusion and very little vacancy ejection at these temperatures. We cannot exclude the possibility of further growth at times many orders of magnitude longer than we can simulate, which may be significant for annealing above 1000K for many hours~\cite{Wang_Acta2024}.
This high temperature mechanism was recently shown in a quasi-2D system by Ishida et al~\cite{Ishida_FusEngDes2025}.
\\

\begin{figure*}
    \centering
    \includegraphics[width=0.75\linewidth]{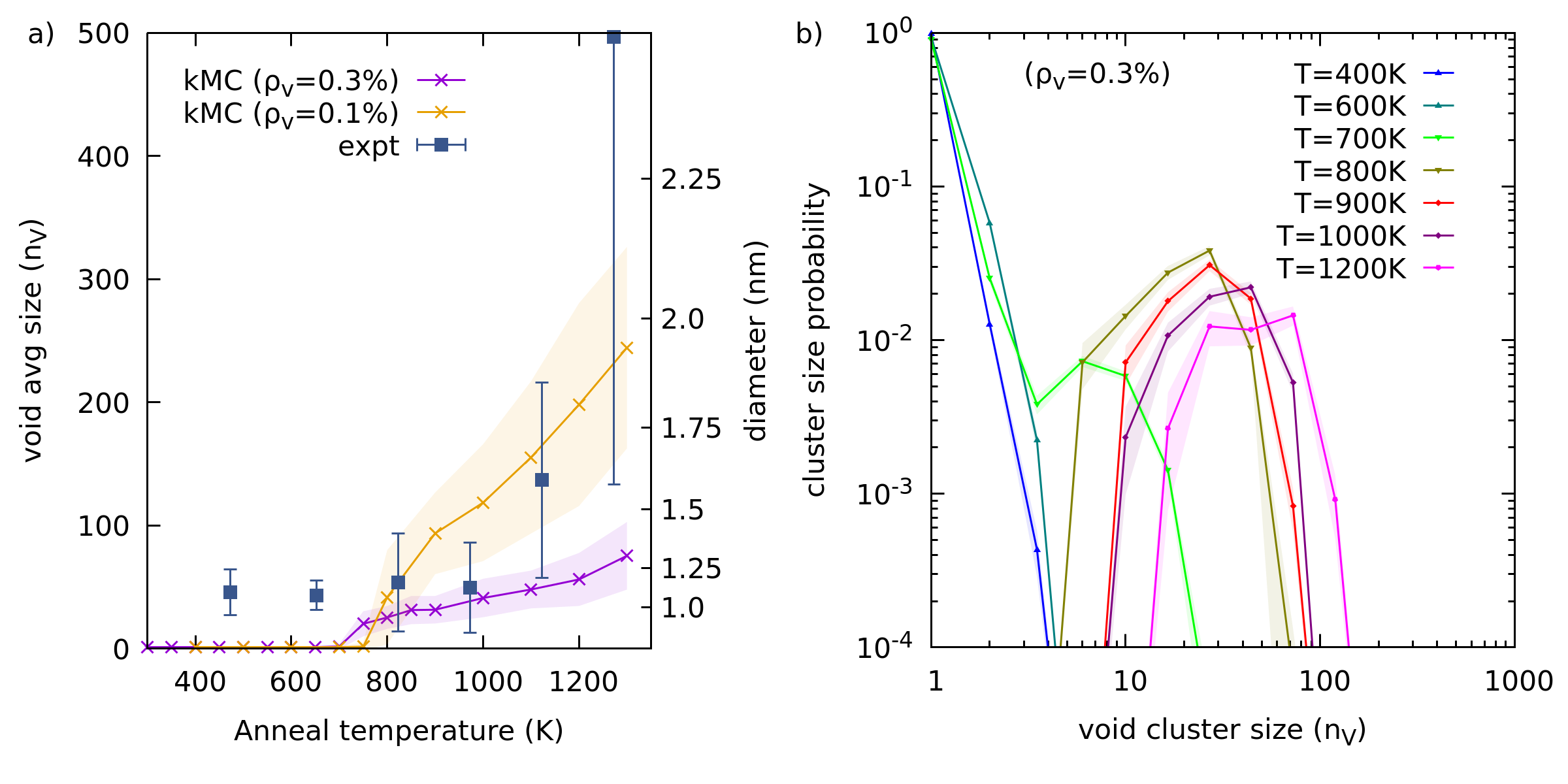}\\
    \caption{
        a) Average cluster size starting with homogeneous vacancy distribution, after 1 hour anneal as a function of temperature. The shaded region is one standard deviation of the void sizes recorded. Experimental points from ref~\cite{Wang_Acta2024} are for heavy-ion irradiation followed by long time annealing.
        b) Vacancy cluster size distribution in tungsten after 1 hour anneal time. 
        Generated from a histogram with log-distributed bins, plotting points at centre of histogram bin with height scaled by bin width. Lines to guide the eye.       
    }
    \label{fig:void_size_histogram}
\end{figure*}

\noindent 
The void cluster size reached after 1 hour annealing is shown in figure \ref{fig:void_size_histogram}a), and compared to an annealing experiment~\cite{Wang_Acta2024}. While our void sizes are broadly in agreement with the experimental data, at lower temperature it is possible that we are able to count vacancy clusters that were almost invisible in the experiment. It is also possible that vacancy clusters, or sponge-like regions which can readily collapse to voids, are initially formed in the ballistic phase of the high-energy ion collisions~\cite{Sand_EPL2016}. At high temperature there may be a further growth phase in tungsten due to Brownian motion of clusters that we are unable to reproduce due to the orders of magnitude more computational time this would require.
\\

\noindent
Figure \ref{fig:void_size_histogram}b) shows cluster size distributions after 1 hour annealing, showing that only the smallest clusters ($n_\mathrm{v} \le 4$) are found at low temperature ($<700$ K), as long-range vacancy migration and aggregation has not yet taken place.
Above 700 K, the vacancies cluster to form ever larger voids. Sensible cluster sizes to consider for a tungsten void study are therefore 15 vacancies for annealing temperatures order 600K--750K, depending on the anneal time and initial vacancy concentration, rising to 60 vacancies for temperatures up to 1200K. 
\\

%%%%%%%%%%%%%%%%%%%%%%%%%%%%%%%%%%%%%%%%%%%%%%%%%%%%%%%%%%%%%%%%%%%%%%

\subsection{Hydrogen Retention in Voids}
\label{sec:voids}

\noindent
For the case of a single hydrogen isotope in a small vacancy cluster, a similar trapping-detrapping rate matrix might be defined as that used in equation \ref{eqn:tridiagG}.
But in a void, we have gas atoms in two distinct thermodynamic states--as surface atoms and as diatoms in the interior. 
Furthermore, gas atoms on one surface site may have a different binding energy to those on a different site, as shown by Hou et al~\cite{Hou_NatMat2019}.
We should therefore have to treat all combinations of surface and diatoms separately to properly describe the kinetics of trapping and detrapping.
\\

\noindent
To simplify the analytic calculation of retention of H atoms in a void, we assert that the H atoms in the diatomic gas in the interior of the bubble are in a dynamic steady state with the H atoms on the surface of the bubble. If there are $n_\mathrm{s}$ surface gas atoms in $N_\mathrm{s}$ equivalent surface sites, the rate of generation of $\mathrm{H}_2$ molecules should be proportional to the number of surface atoms and the occupancy of a neighbour surface site, and be a thermally activated process with an Arrhenius rate factor. The rate of removal of $\mathrm{H}_2$ molecules should be proportional to the number of $\mathrm{H}_2$ molecules, $m$, the square of the fraction of unoccupied surface sites, and also have an Arrhenius rate factor, i.e.
    \begin{eqnarray}
        \label{eqn:surface_to_gas_rates}
        r_{2 \mathrm{H}\rightarrow \mathrm{H}_2} \equiv \tilde{k} &\sim& \frac{1}{2} \frac{n_\mathrm{s}^2}{N_\mathrm{s}} \exp \left[ - \frac{F^{\star} - 2 f_\mathrm{s}(n_\mathrm{s})}{k_B T} \right] \nonumber \\
         r_{\mathrm{H}_2\rightarrow 2\mathrm{H}} \equiv \tilde{p}_m &\sim&  n_{\mathrm{H}_2} \left( 1 - \frac{n_\mathrm{s}}{N_\mathrm{s}} \right)^2 \exp \left[ - \frac{F^{\star} - f_{\mathrm{H}_2}(m)}{k_B T} \right],\nonumber\\
    \end{eqnarray}
where $f_\mathrm{s}=F_\mathrm{s}/n_\mathrm{s}$ is the free energy per surface atom, $f_{\mathrm{H}_2} = F_{\mathrm{H}_2}/m$ is the free energy per $\mathrm{H}_2$ molecule, and $F^{\star}$ is the free energy for the saddle state. 
Note that as we should describe the gas phase with a free energy, we must compare to the free energy of a gas on the surface.
\\

\noindent
We see that equations \ref{eqn:surface_to_gas_rates} give rise to the same tridiagonal form as seen in equation \ref{eqn:tridiagG}, and so, writing the steady state probability of finding the bubble in the state with $m$ gas molecules as $z^{\rm eq}_m$,
    \begin{equation}
        \label{eqn:gas_molecules_eq}
        z^{\rm eq}_m = \frac{ \prod_{i=1}^{m} \tilde{k}/\tilde{p}_m} {1 + \sum_{m'=1}^M \prod_{i=1}^{m'} \tilde{k}/\tilde{p}_{m'}},
    \end{equation}
and so the expected number of gas molecules in a bubble is $\langle n_{\mathrm{H}_2} \rangle = \sum_m m \, z^{\rm eq}_m$. 
Note that because we only need the ratios $k/p_m$, the saddle barrier energy $F^{\star}$ drops out of the steady state expression. The free energies used are described in section \ref{sec:thermodynamics}. We use the free energy of an ideal gas with a steric repulsion term recommended by Tirumala et al~\cite{Tirumala_JPCM2026}. The free energy of a surface gas atom is treated using a vibrational free energy equal to that of the vibrational free energy in the lattice, so that the phononic entropy effect of different surface occupations does not need to be explicitly calculated but the free energy of a gas atom on the surface can be compared to the free energy of a gas molecule. We further use a configurational entropy assuming all surface configurations are degenerate.
Note that these approximations for phonon and configurational free energies make the model easier to parameterize--the success of these approximations is shown below.
\\

\begin{figure*}
    \centering
    \includegraphics[width=0.95\linewidth]{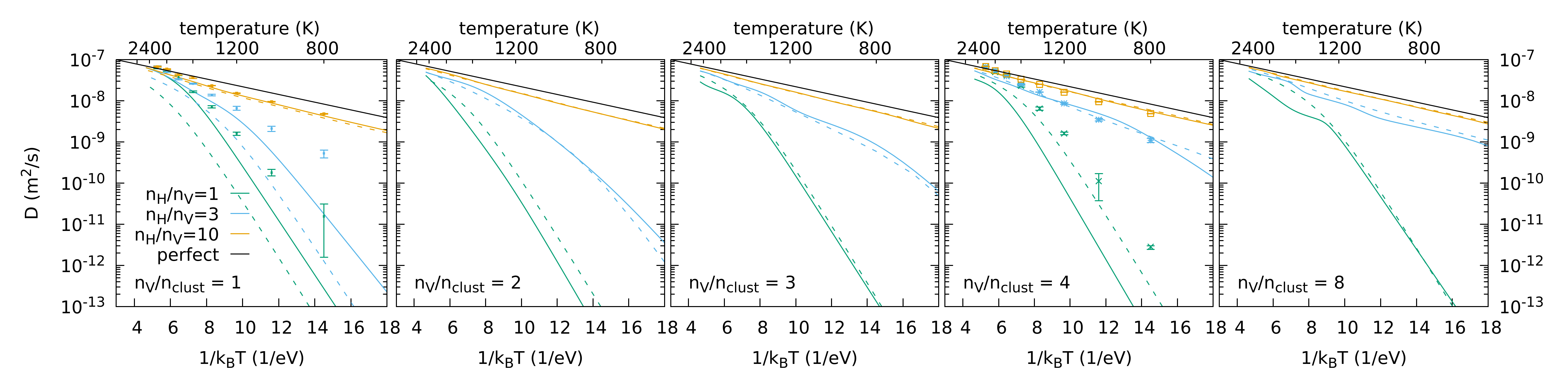}
    \caption{Diffusivity for different sized small vacancy clusters, computed using equation \ref{eqn:steady_state_solution} with the full list of zero-temperature incremental binding energies (solid lines), and using equations \ref{eqn:steady_state_solution} with approximate temperature dependent incremental binding free energies and \ref{eqn:gas_molecules_eq} for the gas molecule count (dashed lines).
    }
    \label{fig:diffusion_explicit}
\end{figure*}

\noindent
We also assert that the H atoms on the surface of the bubble are in a dynamic steady state with the H atoms in mobile interstitial sites. This assumption is tested in MD simulations below. Then the trapping and detrapping rates are as equation \ref{eqn:steady_state_solution}, but with our simple model for the temperature-dependent surface binding energies.
Figure \ref{fig:diffusion_explicit} shows a direct comparison between the two models presented here at small vacancy cluster size- using an explicit calculation of the zero-temperature incremental binding energies and using equation \ref{eqn:steady_state_solution} to compute the expected occupancy of each trap, compared to the approximate but temperature-dependent model for incremental binding free energies. This latter calculation also uses equation \ref{eqn:gas_molecules_eq} for the gas molecule count, but at these small cluster sizes, the gas molecule count is small anyway.
We see that for diffusion through monovacancies ($n_\mathrm{v}/n_\mathrm{clust}=1$), there is a significant difference - and the full binding energy calculation is the better match to the MD simulation data.
But for larger clusters, even divacancies ($n_\mathrm{v}/n_\mathrm{clust}=2$), the difference is much smaller, and it is acceptable to use the approximate model.  
The binding energies used to compute these plots are detailed in section \ref{sec:appendix_binding_energy}.
\\

\subsection{MD simulations of H atoms in voids}

\begin{figure}
    \centering
    \includegraphics[width=0.45\linewidth]{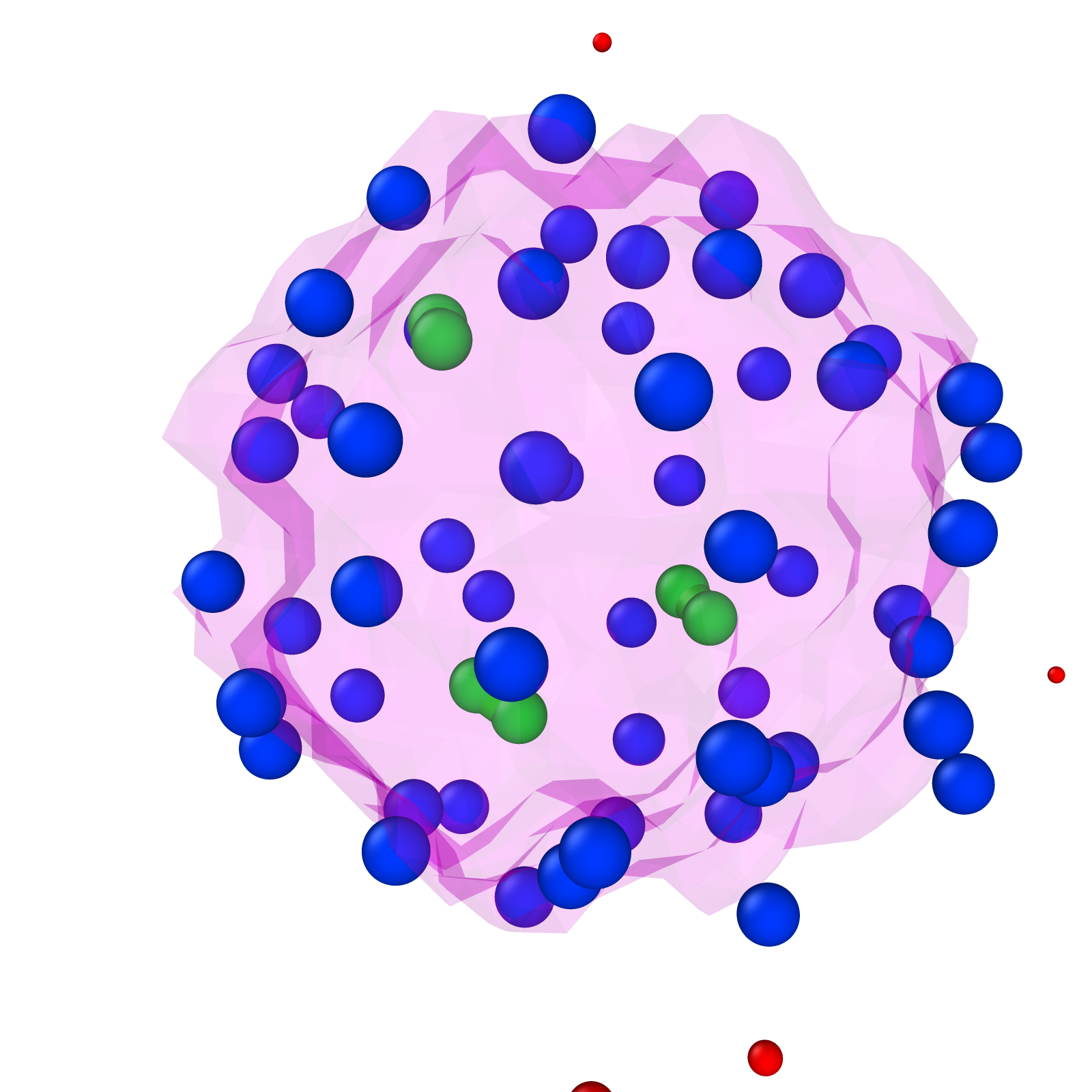}
    \includegraphics[width=0.45\linewidth]{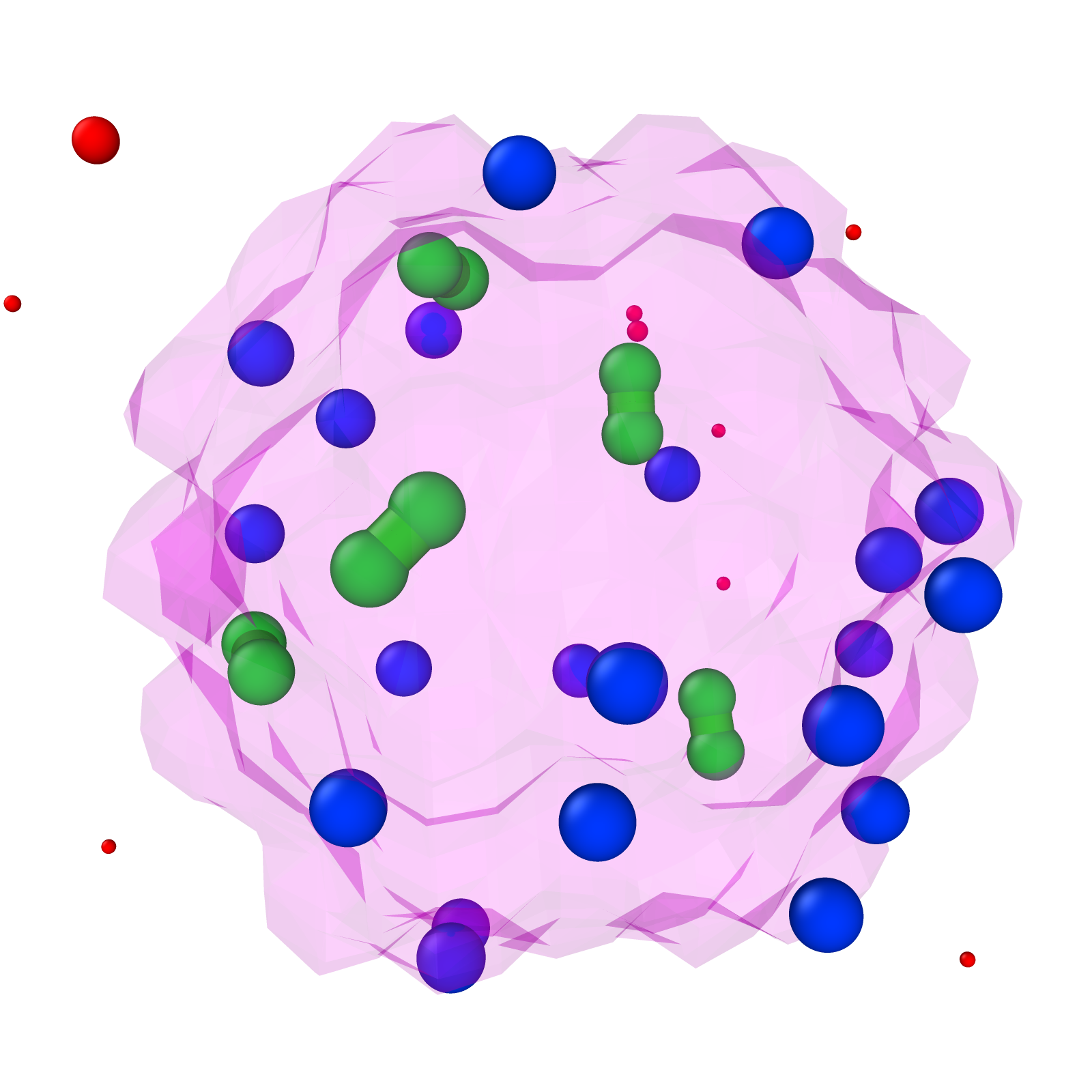}
    \caption{
    Snapshots of MD simulations containing 221k tungsten atoms, with one nano-void containing 60 vacancies (diameter 1.2 nm) and 180 H atoms, $c_\mathrm{H} = 0.08 \,\mathrm{at.} \%$. The void surface is shown in pink using the isosurface method of ~\cite{Mason_PRM2021}, rendered with Ovito~\cite{Stukowski_MSMSE2009}. Surface H atoms are shown in blue, diatomic ${\rm H}_2$ in green, and lattice gas in red.
    Left: the system is thermalized to 800K, and shows 61 H atoms in the void, of which 6 are in ${\rm H}_2$ molecules. Right: thermalized to 1600K, with 31 H atoms in the void of which 10 are in ${\rm H}_2$ molecules. 
    }
    \label{fig:vac60x1.xyz}
\end{figure}

\noindent
We now validate the analytic void retention model, equations \ref{eqn:gas_molecules_eq} and \ref{eqn:steady_state_solution} using direct molecular dynamics simulations in a similar way as in section \ref{sec:MD_simulation} for monovacancies, using the parametrization for tungsten.
We generated initial cells of $48 \times 48 \times 48$ conventional bcc unit cells, and placed a number of vacancy clusters at random. The total monovacancy equivalent count in each cell was fixed at $n_\mathrm{v} = 60$, equivalent to a  concentration of monovacancies $\rho_\mathrm{v}=0.03$ at.\%, but the cluster size varied as $n_\mathrm{v}/n_{\rm clust} \in \{ 4,15,60 \}$. The number of hydrogen atoms was varied to explore different vacancy-hydrogen ratios.
As in the tungsten monovacancy example, we used cycles of annealing then sampling, then slowly reducing the temperature.
Figure \ref{fig:vac60x1.xyz} shows snapshots from the MD simulations, showing H atoms in the lattice, on the surface of the voids, and as $\mathrm{H}_2$ molecules in the interior.
\\

\begin{figure*}[ht]
    \centering
    \includegraphics[width=0.7\linewidth]{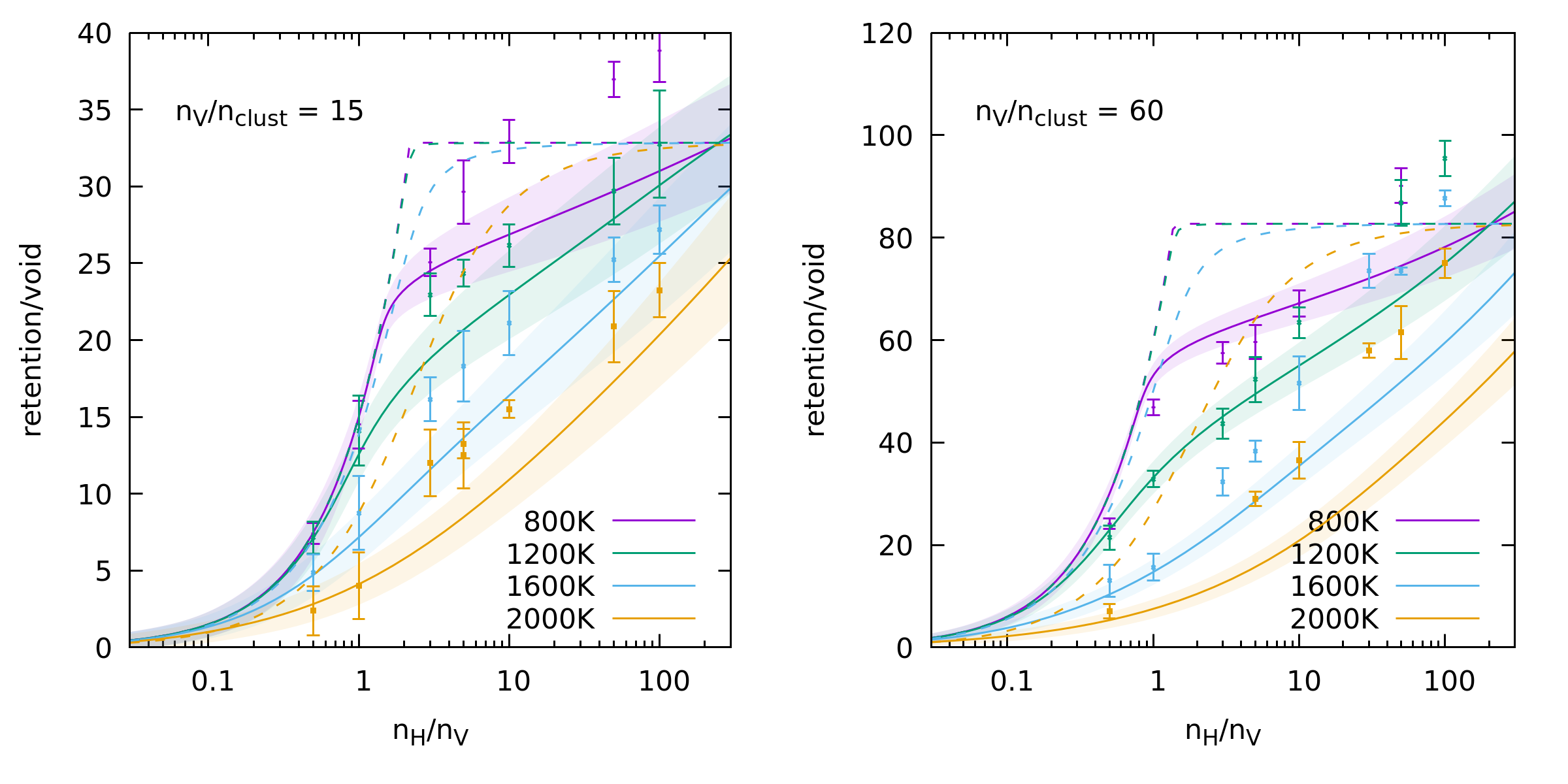}
    \caption{Retention of H atoms (total) in voids computed with MD, compared to analytic curves from our simple model (solid lines). The shaded region is one standard deviation of the retention in the model in the MD system size used.
    (Dashed lines) alternative thermodynamic equilibrium model from ref ~\cite{Zibrov_NME2024}.
    }
    \label{fig:retention_void}
\end{figure*}
\begin{figure}[htb]
    \centering
    \includegraphics[width=0.7\linewidth]{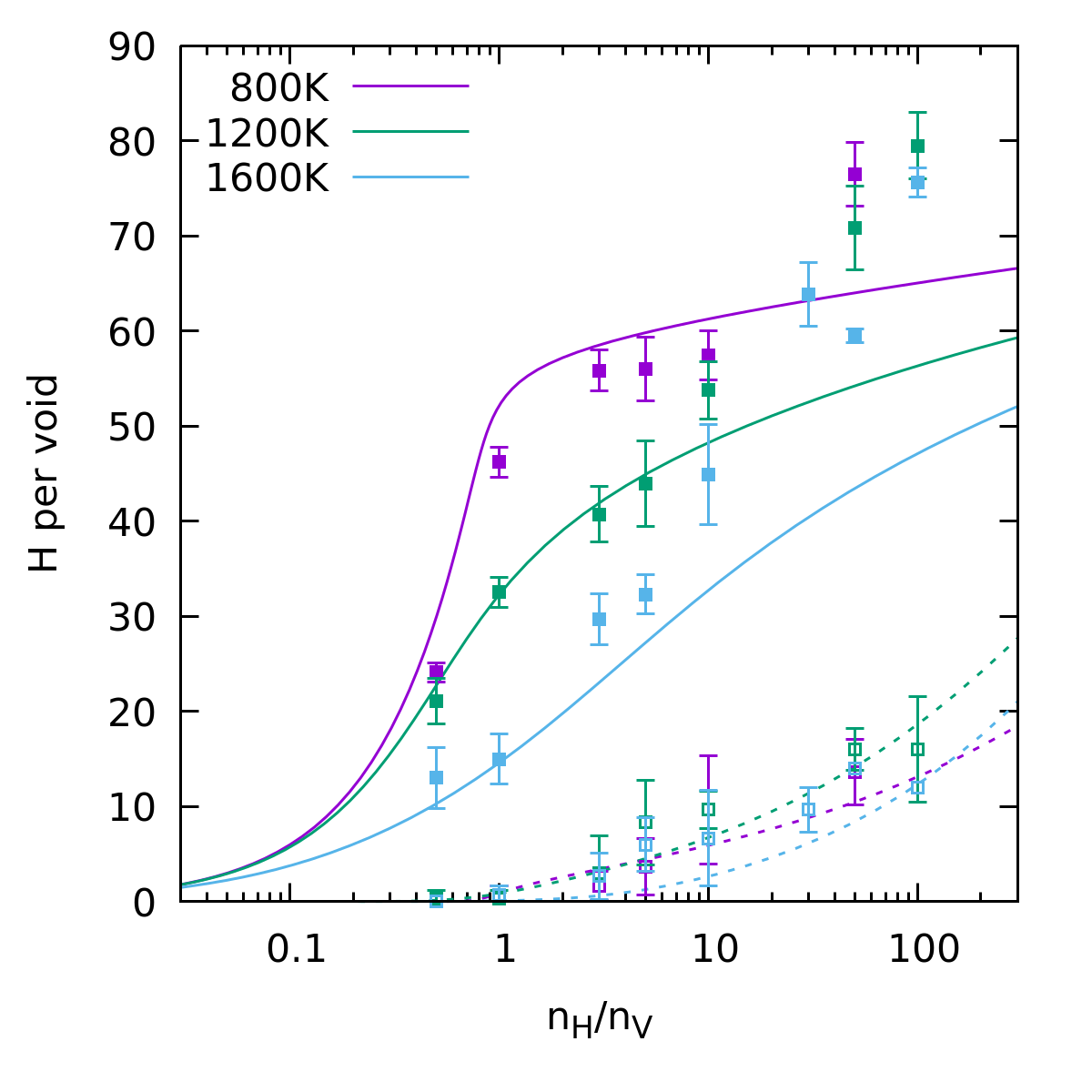}
    \caption{Retention of H atoms on surface in 60-vacancy voids (solid lines) and in the form of $\mathrm{H}_2$ molecules (dashed lines) using our simple model, compared to points generated by direct MD simulation.
    }
    \label{fig:H_surface_and_H2}
\end{figure}

\noindent
Results for the hydrogen retention in the voids in MD simulations as a function of total hydrogen concentration are shown in figure \ref{fig:retention_void}.
As the number of total hydrogen atoms in the box increases, so too does the retention per void. At low temperatures (T $\lessapprox 1000$K) the rise in retention is approximately linear as a function of total H content, until the hydrogen concentration is comparable with the equivalent density of available surface sites. After this point, the rate of rise decreases significantly, and H atoms are found as ${\rm H}_2$ molecules in the void interior. At higher temperatures (T $> 1000$K), the rate of rise starts more slowly, with a less pronounced knee. It is still the case that ${\rm H}_2$ molecules only appear in significant numbers when the hydrogen concentration reaches the equivalent density of available surface sites.
The simple model presented here is a reasonable match to the MD results, showing a quantitative agreement not just for the in total hydrogen retention as a function of temperature, but also the relative proportions of H on surface sites and as $\mathrm{H}_2$ molecules, shown in figure \ref{fig:H_surface_and_H2}.
\\

\noindent
On figure \ref{fig:retention_void} we show an alternative retention model at thermodynamic equilibrium due to Zibrov et al~\cite{Zibrov_NME2024} (their equation 22), parameterized using the same number of surface sites and binding energies. 
The main difference between our model and ref ~\cite{Zibrov_NME2024}, is that they do not have an occupancy-dependent surface binding energy, and do not consider molecular hydrogen in the interior. The model of ref ~\cite{Zibrov_NME2024} does show the qualitative properties of hydrogen retention seen in MD simulation, but with less quantitative accuracy.
\\

\begin{figure*}[htb]
    \centering
    \includegraphics[width=0.9\linewidth]{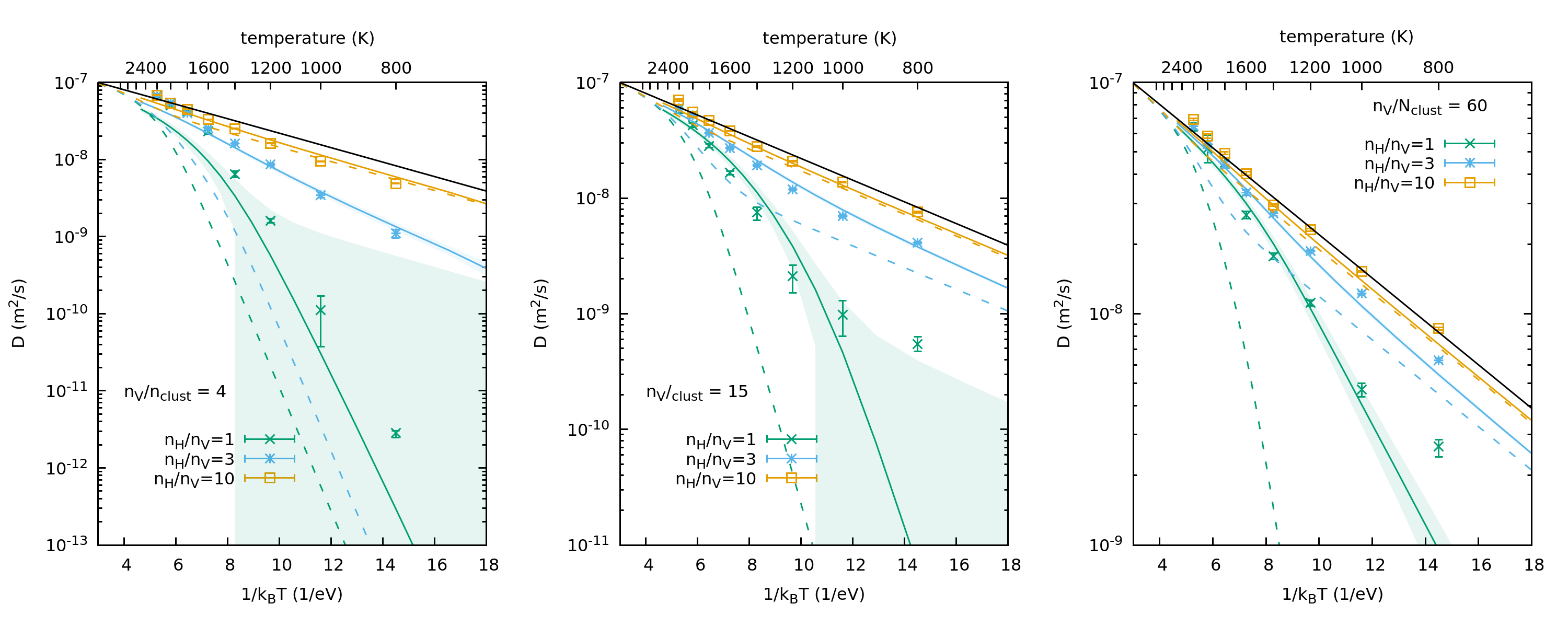}
    \caption{Effective diffusivity computed with MD, compared to analytic curves from our simple model. The shaded regions indicate the expected range of instantaneous diffusivities measured, due to one standard deviation of the expected mobile gas fraction in the MD system size used.
    (Dashed lines) $D_{\rm eff}$ computed using alternative thermodynamic equilibrium model from ref ~\cite{Zibrov_NME2024}.
    }
    \label{fig:diffusion_void}
\end{figure*}

\noindent
Effective diffusivity is plotted in figure \ref{fig:diffusion_void} for three different void sizes, using equation \ref{eqn:EffDiff_mobile}. As there is considerable predicted fluctuation in the occupancy, so too there is a range of mobile gas concentrations, and this leads to a wide range of possible measured instantaneous effective diffusivity, indicated by the shaded region in the plots. 
We see that our simple model is a good match to the direct MD simulation results, but not perfect. 
The discrepancy is clearest for low occupancy at 800K, where we predict a lower diffusivity than measured. Comparing back to figure \ref{fig:retention_void}, we see that our simple model is slightly overestimating retention, and so underestimating the mobile gas content. The diffusivity is very sensitive to small changes in mobile gas content in these calculations--there may be only one or two mobile gas atoms in the MD simulation box.
In principle, we could adjust the parameters of our simple model to improve the fit to the observed MD result. Adjusting the number of surface sites is the easiest way to do this, as it does not significantly affect the balance between surface and molecular H atoms. However, this would need to be done for each void size, and as a range of void sizes and shapes will be present in a real-life situation, we elect to present a fair but easily transferable model rather than adjusting parameters to make a perfect fit.
\\

\noindent
Also marked on figure \ref{fig:diffusion_void} is an effective diffusivity derived using equation \ref{eqn:EffDiff_total} but with the mobile fraction computed using a retention from ref ~\cite{Zibrov_NME2024}. We can see that even though the retention in this model is qualitatively correct, the sensitivity of the diffusivity to the mobile gas fraction means that the diffusion estimate in this model is much further from the MD results.
\\

\subsection{Hydrogen retention and diffusion in post-irradiation annealed void microstructures}
\label{sec:PIA_voids}
 
\begin{figure*}
    \centering
    \includegraphics[width=0.7\linewidth]{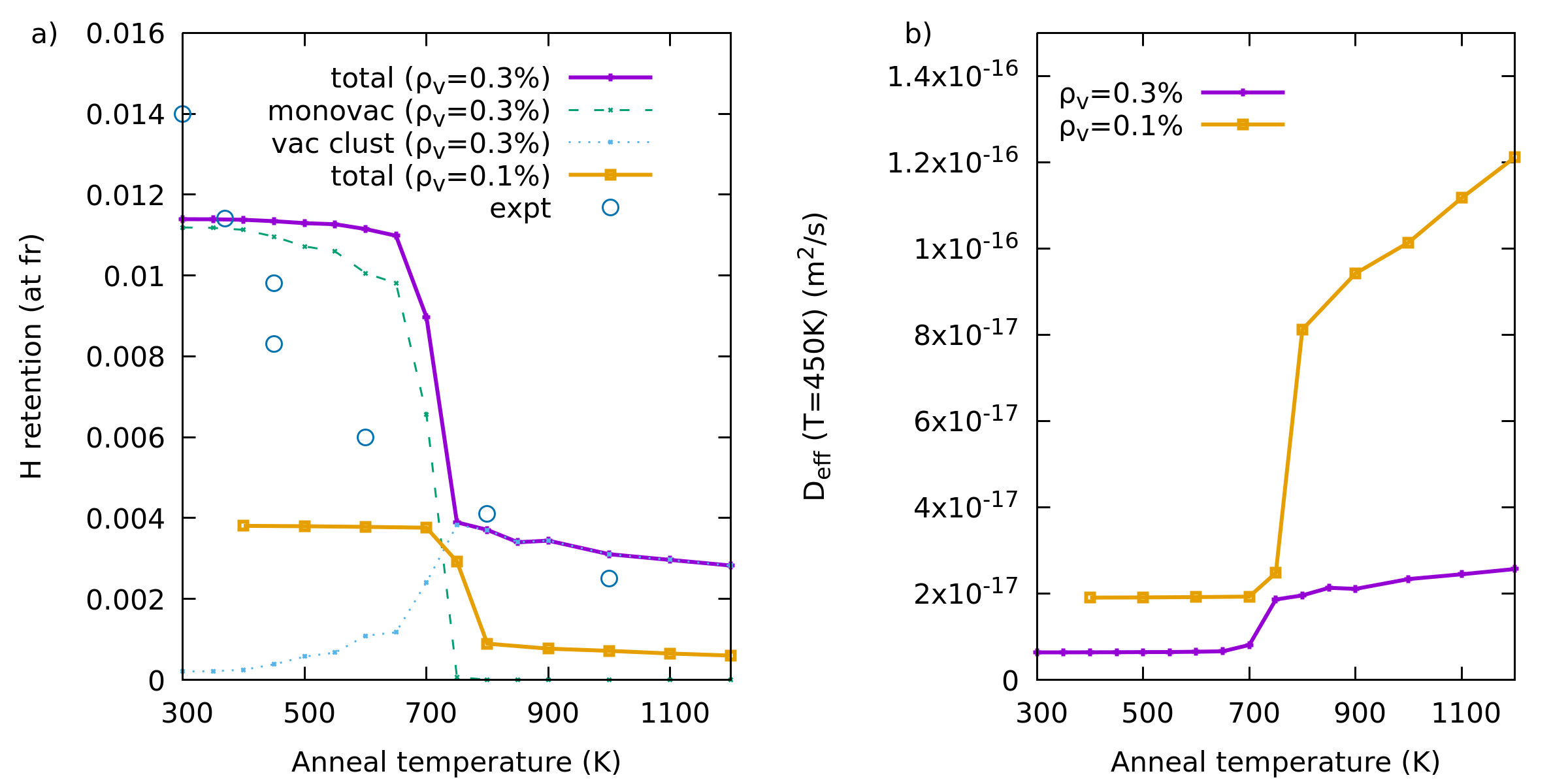}
    \caption{
    a) H retention after 1 hour anneal time, assuming plasma loading gives mobile content $x=10^{-8}$ at \% and loading at 450 K. Experimental data points are D retention after heavy ion irradiation damage to 0.35 dpa at different irradiation temperatures, from ref~\cite{Markelj_PhysScripta2022}.
    b) H effective diffusivity at $x=10^{-8}$ at \% and 450 K after annealing for 1 hour. 
    }
    \label{fig:void_retention_diffusion}
\end{figure*}

\noindent
The final calculation in this section combines the void size histograms (figure \ref{fig:void_size_histogram}) with the now validated form for retention and diffusion in voids to make predictions for more realistic void cluster size distributions. Recall that our void cluster sizes assume that irradiation has made a homogeneous monovacancy content at atomic fraction $\rho_\mathrm{v}$, which is then post-irradiation annealed for 1 hour. 
We assume that the irradiated layer is a few microns only on a slab of material order 1 mm thickness.
We consider a simple loading model, namely a steady plasma flux is maintained until the mobile concentration is steady at through the irradiated region, and the vacancies and voids are in a saturated steady state with the mobile population.
We use a typical  mobile gas fraction at the surface for a plasma-loading experiment of $x=10^{-8}$ at \% using eqn 20 in ref \cite{hodille2018}.
The temperature of the irradiated tungsten is kept at 450K through the gas loading.
\\

\noindent
The predicted retention is shown in figure \ref{fig:void_retention_diffusion}a). 
Below 750K, we predict little void growth after 1 h annealing, so the retention is nearly all in monovacancies.
Over 750K, the clustering starts and the number of monovacancies drops rapidly, and retention is in void clusters.
As the maximum retention in a void cluster scales with the surface area, coalescing vacancies in this way markedly reduces the retention, so we see a sharp drop in H retention above 750K.
As the temperature increases further, the average void size increases slowly, with a corresponding slow decrease in saturated H retention.
In figure \ref{fig:void_retention_diffusion}a) we also plot experimental data points for a similar but not identical scenario--irradiation at different temperatures rather than post-irradiation annealing~\cite{Markelj_PhysScripta2022}.
The magnitude of the saturated retention observed in experiments is the same order of magnitude as our simple models.
In particular we note that assuming $\rho_\mathrm{v} = 0.3$ at \%, as seen in large scale overlapping cascade simulations~\cite{Boleininger_SciRep2023}, gives a very good match to the experimentally observed saturated retention at low temperature, as has been noted previously~\cite{Mason_PRL2020}.
\\
\noindent
At intermediate annealing temperatures, {$T$=300--750\,K}, interstitial loops produced during irradiation have some mobility, with a range of thermal activation barriers. In experiment, therefore, we would expect to see a slow drop of vacancy content in this annealing temperature range. This mechanism for vacancy count reduction is missing from our simple annealing model, as we have ignored interstitials completely. Hence our estimate for retention with fixed $\rho_\mathrm{v} = 0.3$ at \% is too high.
A new result here is that the drop in retention at irradiation/annealing temperatures of 750K is well described by our simple model, which is consistent with interstitial loops rapidly being lost to other sinks and not annihilating on voids. 
\\

\noindent
The predicted effective diffusivity during gas loading is shown in figure \ref{fig:void_retention_diffusion}b).
While we do not know of any experimental data to confirm this, our prediction is that the diffusivity rapidly increases at irradiation/annealing temperatures of 750K, at the point of void coalescence. 
This is for the same reason as the retention drop--coalescence of vacancies reduces the number of high binding energy surface sites, and so the fraction of mobile gas atoms is higher with voids than with monovacancies.
\\

%%%%%%%%%%%%%%%%%%%%%%%%%%%%%%%%%%%%%%%%%%%%%%%%%%%%%%%%%%%%%%%%%%%%%%

\section{Conclusion}

\noindent 
In this work we have performed MD simulations of hydrogen diffusing in tungsten and vanadium containing vacancies at high concentrations comparable with saturated ion-irradiation damage. We considered the diffusion and retention in monovacancies, small vacancy clusters, and large voids.
This we compared to analytic expressions for the diffusivity and retention in multi-occupancy traps.
We show that with a simple model for the free energy of hydrogen atoms in a void, we can reproduce both retention and diffusivity. This is not possible to achieve with a single-occupancy trap model across the range of input concentrations considered here.
\\

\noindent
To determine the appropriate void size for ion-irradiated materials, we used rejection kinetic Monte Carlo to find reasonable void sizes using similar potentials and assumptions.
This short study was essential to complete the validation of this work, as if carbon or other impurities were needed to produce voids of the observed sizes in tungsten, they should also be present in the hydrogen diffusion-retention study.
Notably we found that voids grow with a nucleation-and-growth mechanism, with representative bcc metals showing an initial rapid clustering followed by Brownian motion of small clusters. Tungsten is exceptional in that it has very low binding energy for small clusters, and instead very slowly nucleates large clusters, which then show little (but not zero) Brownian motion. 
Modelling the evolution of complex irradiation-induced microstructure including dislocation loops at and beyond stage III recovery (long-range vacancy diffusion) is beyond the scope of this paper, as we are focussing on hydrogen retention.
\\

\noindent
The results we have presented here are for protium in bcc metals. 
The analytic expressions can be adjusted easily to account for deuterium or tritium. The MD simulations do not include zero-point energy effects, and in the tungsten case use empirical potentials, so caution should be used in over-interpreting quantitative values.
\\

\noindent
We considered the effect of making different physical approximations for the detrapping rate geometric/entropic prefactor $g_i'$, and concluded that it was difficult to recommend one model which would be best for both tungsten and vanadium. We recommend using a prefactor which will give thermodynamic equilibrium, including configurational entropic effects, largely because it is easier to compute than including phononic entropic effects. This lack of a definitive result could be the subject of a larger study across a wider range of materials, but is outside of the scope of this paper.
\\

\noindent
 The fit of the MD simulation data to the steady state generalised Oriani expression for monovacancies is very good--here we had full data for the binding energies of all the trapped and untrapped configurations to put into equation \ref{eqn:tridiagG}. With void simulations calculating every configuration at low temperature is prohibitive, so we used a simple steady-state model based on matching the chemical potential of hydrogen in lattice sites, surface sites and as a diatomic gas, and computed the retention analytically. Here we still found a good agreement with the MD results--both in the retention of hydrogen on the surface of a void and as ${\rm H}_2$ gas in the void interior, and also in the measured effective diffusion constant.
 \\

\noindent 
This work has validated analytic expressions for the retention and diffusivity of gas in vacancy-type traps characteristic of irradiation damage in structural metals, and we were able to make a forward prediction from first principles for the retention and diffusion of hydrogen in a simple model for irradiated tungsten subjected to post-irradiation annealing.
We conclude that modelling diffusion and retention with multi-occupancy traps is not just practical, but essential when considering larger defects.
\\

%%%%%%%%%%%%%%%%%%%%%%%%%%%%%%%%%%%%%%%%%%%%%%%%%%%%%%%%%%%%%%%%%%%%%%

\section*{Acknowledgements} \label{sec:acknowledgements}

\noindent
The authors would like to thank Duc Nguyen-Manh for stimulating discussions in the development and validation of our model.
\\

\noindent
This work has been carried out within the framework of the EUROfusion Consortium, funded by the European Union via the Euratom Research and Training Programme (Grant Agreement No 101052200 — EUROfusion) and from the EPSRC [grant number EP/W006839/1].  Views and opinions expressed are however those of the author(s) only and do not necessarily reflect those of the European Union or the European Commission. Neither the European Union nor the European Commission can be held responsible for them. This work has been (part-) funded by the EPSRC Energy Programme [grant number EP/W006839/1].  
PS would like to acknowledge the Department for Science, Innovation and Technology's International Science Partnership Fund (ISPF). 
Some of this work is part of the UKAEA’s Lithium Breeding Tritium Innovation (LIBRTI) programme, funded by the Department of Energy Security and Net Zero, as announced by the UK Government in October 2023. \\

\noindent
HPC time was provided by EUROfusion on the Leonardo machine, and by the Cambridge Service for Data Driven Discovery (CSD3) and associated support services provided by the University of Cambridge Research Computing Services (www.csd3.cam.ac.uk). Some of this work used the ARCHER2 UK National Supercomputing Service (https://www.archer2.ac.uk)~\cite{archer2}.
\\

\noindent
To obtain further information on the data and models underlying this paper please contact PublicationsManager@ukaea.uk.  \\

%%%%%%%%%%%%%%%%%%%%%%%%%%%%%%%%%%%%%%%%%%%%%%%%%%%%%%%%%%%%%%%%%%%
%%%%%%%%%%%%%%%%%%%%%%%%%%%%%%%%%%%%%%%%%%%%%%%%%%%%%%%%%%%%%%%%%%%
%%%%%%%%%%%%%%%%%%%%%%%%%%%%%%%%%%%%%%%%%%%%%%%%%%%%%%%%%%%%%%%%%%%

\section*{Data availability}\label{sec:data_availability}

\noindent
The code, scripts and data used in this paper will be made available on acceptance. \\

%%%%%%%%%%%%%%%%%%%%%%%%%%%%%%%%%%%%%%%%%%%%%%%%%%%%%%%%%%%%%%%%%%%
%%%%%%%%%%%%%%%%%%%%%%%%%%%%%%%%%%%%%%%%%%%%%%%%%%%%%%%%%%%%%%%%%%%
%%%%%%%%%%%%%%%%%%%%%%%%%%%%%%%%%%%%%%%%%%%%%%%%%%%%%%%%%%%%%%%%%%%

\section*{CRediT}\label{sec:CRediT}

\noindent
{\bf D. Mason} Conceptualisation, Methodology, Software, Validation, Formal Analysis, Writing, Supervision,
{\bf S. Kaur} Formal Analysis, Validation,
{\bf S. Tirumala} Investigation,
{\bf P. Srinivasan} Investigation,
{\bf V. Jantunen} Methodology, Investigation,
{\bf M. Boleininger} Conceptualisation, Supervision.

%%%%%%%%%%%%%%%%%%%%%%%%%%%%%%%%%%%%%%%%%%%%%%%%%%%%%%%%%%%%%%%%%%%

\section*{Declarations}\label{sec:declarations}

\noindent
The authors have no conflicts of interest to declare. 

%%%%%%%%%%%%%%%%%%%%%%%%%%%%%%%%%%%%%%%%%%%%%%%%%%%%%%%%%%%%%%%%%%%
%%%%%%%%%%%%%%%%%%%%%%%%%%%%%%%%%%%%%%%%%%%%%%%%%%%%%%%%%%%%%%%%%%%
%%%%%%%%%%%%%%%%%%%%%%%%%%%%%%%%%%%%%%%%%%%%%%%%%%%%%%%%%%%%%%%%%%%

\bibliographystyle{unsrt}
\bibliography{references}

\section*{Appendix}

\label{sec:thermodynamics}
\noindent
In this appendix, we derive a simple model for the free energy of hydrogenic gases in a bubble.
\\

\noindent
In a non-hydride forming metal containing nanovoids, hydrogen gas can be found in interstitial lattice sites, at the surface of bubbles, or as a gas inside bubbles.
Given a fixed number of H atoms total, we want to find the best distribution of hydrogen between these three subsystems, in the sense that free energy is minimised. 
Each subsystem has different degrees of freedom--the gas in the lattice has a large number of sites to explore, while the gas on the void surface has fewer sites but has a binding energy. The diatomic gas in the bubbles has translational and rotational degrees of freedom.
Therefore the free energy of each subsystem varies differently as a function of temperature, and to solve for the distribution of hydrogen needs a model for the partition function for each subsystem, and not just a single zero-temperature binding energy.
\\

\subsection{Free energy of atomic hydrogen in lattice}

The free energy of a hydrogen atom in the lattice can be found by considering it to be a quantum harmonic oscillator, with has free energy 
    \begin{equation}
        F^{\rm vib}_{\rm L} = \frac{ 3  \hbar \omega_{\rm L}}{2} + 3 k_B T \log\left( 1 - \exp\left( - \frac{\hbar \omega_{\rm L} }{k_B T} \right) \right).
    \end{equation}
In the classical limit and neglecting zero point energy, $F^{\rm vib}_{\rm L} \approx 3  k_B T \log\left( \frac{\hbar \omega_{\rm L} }{k_B T} \right)$.
The lattice hydrogen also has a formation energy, $E^{\rm f}_{\rm L}$, defined as the energy per atom required to split the ${\rm H}_2$ molecule bond and place it into a lattice site. For a non-hydride forming metal, $E^{\rm f}_{\rm L}>0$.
In a bcc metal, the H atom sits at tetrahedral sites, of which there are $\gamma=6$ per host atom.
We do not consider interactions between hydrogen atoms on neighbouring lattice sites. This is a reasonable approximation for low concentrations in tungsten~\cite{Becquart_JNM2009b}.
We also need to consider the configurational entropy of placing atoms on discrete lattice sites. If there are $n_{\rm L}$ lattice gas atoms with $N$ host atom sites, the total Helmholtz free energy of the lattice gas is 
    \begin{equation}
        \label{eqn:freeEnergyLatticeGas}
        F_{\rm L}( n_{\rm L};T) = n_{\rm L} E^{\rm f}_{\rm L} + n_{\rm L} F^{\rm vib}_{\rm L} - k_B T \log\left( \frac{ (N\gamma)! }{(N\gamma-n_{\rm L})! n_{\rm L}! } \right).
    \end{equation}
The chemical potential associated with this free energy is
    \begin{eqnarray}
        \mu_{\rm H} &\equiv& \frac{\partial F_{\rm L}}{\partial n_{\rm L}} =  E^{\rm f}_{\rm L} + F^{\rm vib}_{\rm L}  + k_B T \log \left( \frac{n_{\rm L}} {N\gamma-n_{\rm L}} \right) \nonumber\\
        &=& E^{\rm f}_{\rm L} + F^{\rm vib}_{\rm L}  + k_B T \log \left( \frac{x} {\gamma-x} \right),
    \end{eqnarray} 
where $x$ is the atomic fraction of lattice gas.

\subsection{Free energy of diatomic hydrogen in gas}

\noindent
The Gibbs free energy of $n$ ideal diatomic gas molecules is
    \begin{equation}
        G_{\rm dia} = n E^{\rm f}_{{\rm H}_2} + p V - k_B T \log Z_{\rm dia},
    \end{equation}
where the partition function of the ideal diatomic gas is separable into its component degrees of freedom per molecule,
    \begin{equation}
        Z_{\rm{dia}} = \frac{ \left( z_{\rm{trans}} z_{\rm{rot}} z_{\rm{vib}} z_{\rm{elec}}  z_{\rm{nucl}} \right)^n} {n!}.
    \end{equation}
We will consider the component degrees of freedom separately. We will ignore vibrational, electronic and nuclear spin degrees of freedom, as the vibrational and electronic degrees of freedom are important at very high temperature and the nuclear at very low temperature compared to the operating temperatures we are interested in.
We will take the binding energy of the diatomic molecule, $E^{\rm f}_{{\rm H}_2}$, to be zero by convention.
\\

The translational degrees of freedom for the diatomic gas will be the \textit{thermal wavelength}, with the masses of the hydrogen isotopes comprising the diatom $m_1$ and $m_2$.
     \begin{equation}
        Z_{\rm{trans}} = \frac{V^n}{n!} \left( \frac{ 2 \pi (m_1+m_2) k_B T }{h^2 } \right)^{3n/2},
    \end{equation}

We will use the rigid rotor-harmonic oscillator approximation to separate the vibrational and rotational modes.
For a heteronuclear diatomic molecule, the rotational partition function is given by a sum over the rotation eigenstates $J = 0,1,2\ldots$,
    \begin{eqnarray}
        \label{eqn:rotational}
        z^{\rm{het}}_{\rm{rot}} &=& \sum_{J=0}^{\infty} \exp\left( -\frac{e_J}{k_B T} \right) \nonumber \\
        &=& \sum_{J=0}^{\infty} (2J+1) \exp\left( -\frac{\Theta_{\rm{rot}} J(J+1) }{k_B T} \right)
    \end{eqnarray}
where the rotational temperature $\Theta_{\rm{rot}}$ is defined by 
    \begin{equation}
        \Theta_{\rm{rot}} = \frac{h^2  }{8 \pi^2 \tilde{m} r^2 k_B},
    \end{equation}
where $\tilde{m} = m_1 m_2 / (m_1+m_2)$ is the reduced mass and $r$ the atom's bond length.
This temperature constant for hydrogen isotopes can be found using the value for the ${\rm H}_2$ molecule, $\Theta_{\rm{rot}}^{{\rm H}_2} = 59.3$K.
For a diatomic molecule with reduced mass $\tilde{m}$,
    \begin{equation}
        \Theta_{\rm{rot}} = \frac{m_{H}}{2 \tilde{m}} \Theta_{\rm{rot}}^{{\rm H}_2} 
    \end{equation}

\begin{comment}    
The expression in equation \ref{eqn:rotational} is difficult to work with except in its limits. In the high temperature limit $\Theta_{\rm{rot}} \ll T$,
    \begin{equation}
        z_{\rm{rot}}^{\rm{het}} \approx \frac{T}{\Theta_{\rm{rot}}}.
    \end{equation}

For a \emph{homonuclear} diatomic molecule, the situation is more complicated as the wavefunction must be antisymmetric under the interchange of nuclei. This is of great importance at low temperatures, but at high temperatures (over room temperature), we have so many total rotational degrees of freedom active that this difference can be averaged out, and we end up with the simple expression
    \begin{equation}
        z_{\rm{rot}}^{\rm{homo}} \approx \frac{T}{2 \Theta_{\rm{rot}}}.
    \end{equation}
\end{comment}

We approximate equation \ref{eqn:rotational} the Euler-MacLaurin expansion,
    \begin{equation}
        z_{\rm{rot}}^{\rm{dia}} = \frac{T}{\sigma \Theta_{\rm{rot}}}\left( 1 + \frac{1}{3}\frac{\Theta_{\rm{rot}}}{T} + \frac{1}{15} \left( \frac{\Theta_{\rm{rot}}}{T} \right)^2 + \ldots \right),
    \end{equation}
where the symmetry number $\sigma = 1$ for heteronuclear molecules and $\sigma = 2$ for homonuclear molecules.

The total partition function for the diatomic hydrogen isotope molecule between room temperature and 2000K is approximately
    \begin{eqnarray}
        Z_{\rm{dia}} &\approx& \frac{ \left( z_{\rm{trans}} z_{\rm{rot}} \right)^n} {n!} , 
    \end{eqnarray}
and so the Gibbs free energy of the diatomic gas is
    \begin{eqnarray}
        \label{eqn:GibbsFreeEnergyDiatomicGas}
        &&G_{\rm{dia}}(n_{{\rm H}_2},V,T)  =   p V - k_B T \log Z_{\rm{dia}}    \nonumber \\
        \quad \quad &\approx& - n_{{\rm H}_2} k_B T \log \left( \frac{k_B T^2}{p \, \sigma \Theta_{\rm{rot}}^{{\rm H}_2}} \left( \frac{2 \pi m_{\rm H} \left(m_1+m_2\right)^2 k_B T }{h^2 \, m_1 m_2} \right)^{3/2} \right)  \nonumber \\
        \quad \quad &=& n_{{\rm H}_2} k_B T \log \left( \frac{p}{p^*} \right)   \nonumber   \\
        \quad \quad &=& n_{{\rm H}_2} k_B T \log \left( \frac{n_{{\rm H}_2} k_B T}{p^* V} \right)
    \end{eqnarray}

% The chemical potential is 
%     \begin{eqnarray}
%         \label{eqn:ChempotDiatomicGas}
%         \mu_{\rm{dia}} &=& \left( \frac{\partial G_{\rm{dia}} }{\partial n} \right)_{p,T} \nonumber\\
%         &=&  k_B T \log \left( \frac{p}{p^*} \right),  
%         %&=&  E^f_{{\rm H}_2} + k_B T \log \left[ \frac{p}{p_0(T)  \times \Phi(m_1,m_2) }
%         %\frac{ \sigma\left(\frac{2 \mu}{m_H}\right)^2  }{\left(\frac{m_1+m_2}{2 m_H} \right)^{3/2}}    \right],
%     \end{eqnarray}
where we can identify the characteristic pressure, $p^*$, as the product $p^* = p_0(T)  \times \Phi(m_1,m_2) $, where
$p_0(T)$ is a temperature dependent constant defined by the ${\rm H}_2$ molecule,
    \begin{equation}
        \label{eqn:pressureConst}
        p_0(T) = \frac{k_B T^2}{\Theta_{\rm{rot}}^{{\rm H}_2} } 
        \left( \frac{2 \pi m_{\rm H}  k_B T }{h^2 } \right)^{3/2},
    \end{equation}
and $\Phi(m_1,m_2)$ is an isotope dependent correction,
    \begin{equation}
        \label{eqn:gasConst}
       \Phi(m_1,m_2) = \frac{1}{\sigma} \,  \frac{\left( m_1 + m_2 \right)^3 }{\left( m_1 m_2 \right)^{3/2}} .
    \end{equation}
%Note that the symmetry number is a quantum mechanical effect. For a pure classical model, replace $\sigma \rightarrow 1$.

Note that $p/p_0 \sim T^{-5/2}$. From this it follows that internal energy of the classical diatomic gas, $U$, is
    \begin{equation}
        U \equiv - \frac{ \partial \log Z }{\partial \beta} = \frac{5}{2} n_{{\rm H}_2} k_B T.
    \end{equation} 
Note that as we have chosen the ${\rm H}_2$ bond energy to be the zero point of energy, this internal energy is all kinetic.

\subsection{Steric effects}

\noindent
There is short-range repulsion between gas molecules. If at high density, the energy is of the Lennard-Jones 6-12 pairwise type $E^{\rm st} \sim \varepsilon/d^{12}$, so the extra energy due to steric effects is order
    \begin{equation}
        E^{\rm st} = \frac{\varepsilon}{2} \left( \frac{n_{{\rm H}_2}}{\tilde{V}} \right)^4 n_{{\rm H}_2}.
    \end{equation}
The factor $1/2$ is for double-counting pairs of ${\rm H}_2$ molecules, and $\langle d\rangle^3\sim \frac{\tilde{V}}{n_{{\rm H}_2}}$.
We write the accessible volume of the bubble interior as $\tilde{V} \le V$, as there may be some excluded volume near the surface of the bubble where ${\rm H}_2$ atoms can not exist.
We can parameterize this exclusion effect with a single excluded surface layer thickness, $\delta r$, to give
    \begin{equation}
        \tilde{V} = \frac{\pi}{6}\left( \left( \frac{6 V}{\pi} \right)^{1/3} - 2 \delta r \right)^{3}.
    \end{equation}
Including this excluded surface layer means that diatomic gas does not form in very small vacancy clusters, in line with zero temperature observations using density functional theory~\cite{Hou_NatMat2019}.
The steric effect is negligible except in over-pressurised bubbles. The constant $\varepsilon$ is a single parameter characterising the interaction strength with unit dimensions $[E][V^4]$.
\\

\noindent
Including this term into the 
%internal energy gives 
%    \begin{equation}
%        U_{\rm st} = \frac{\varepsilon}{2} \frac{n_{{\rm H}_2}^5}{V^4} + \frac{5}{2} n_{{\rm H}_2} k_B T,
%    \end{equation}
%and including it into the 
gas free energy gives a simple non-ideal gas law for bubbles,
    \begin{equation}
        G_{\rm dia,st}(n_{{\rm H}_2};V,T) = E^{\rm st}\left( n_{{\rm H}_2} ; V \right) + n_{{\rm H}_2} k_B T \log \left( \frac{n_{{\rm H}_2} k_B T}{p^* V} \right).
    \end{equation}  
\\

\subsection{Surface binding}

Let us assume there are surface binding sites associated with a void/bubble.
The number of these sites, $N_{\rm s}$, will increase with surface area. We know that a monovacancy in bcc has $\beta \gtrapprox 6$ surface sites, so say
    \begin{equation}
        N_{\rm s}^{\rm bcc}(V) = \beta \left( \frac{2 V}{a_0^3} \right)^{2/3},
    \end{equation}
with $a_0$ the bcc lattice parameter.
The simplest model for the binding energy is to say that the binding energy is constant at low occupation, falling to zero when the number of surface sites is exceeded. A good empirical fit to the monovacancy binding energy is quartic,
    \begin{equation}
        E^{\rm b}_{\rm s}(n_{\rm s}) = \alpha \left( 1 -  \left( \frac{ n_{\rm s} }{ N_{\rm s}} \right)^4 \right).
    \end{equation}
Note this model has an adjustable energy parameter, $\alpha$, which itself may be a function of void volume, though we treat it as a constant fit to the monovacancy for this work. This model gives a formation energy for $n_{\rm s}$ hydrogen atoms on the surface of a void,
    \begin{eqnarray}
        E^{\rm f}_{\rm s}(n_{\rm s}) &=& E^{\rm f}_{\rm s}(n_{\rm s}-1) + F^{\rm f}_{\rm L} - E^{\rm b}_{\rm s}(n_{\rm s})    \nonumber\\    
        &=& n_{\rm s} F^{\rm f}_{\rm L} - \sum_{n=1}^{n_{\rm s}} E^{\rm b}_{\rm s}(n)   \nonumber\\ 
        &=& n_{\rm s} (F^{\rm f}_{\rm L}-\alpha) + \alpha \, \frac{ n_{\rm s}(n_{\rm s}+1)(2n_{\rm s}+1) }{6 N_{\rm s}^2 } \nonumber\\
        &\approx& n_{\rm s} (F^{\rm f}_{\rm L}-\alpha) + \alpha \, \frac{ n_{\rm s}^3 }{3 N_{\rm s}^2 } .
    \end{eqnarray}

The free energy of the surface atom can be found by considering it to be a quantum harmonic oscillator; we also need to consider the configurational entropy of placing $n_{\rm s}$ atoms into $N_{\rm s}$ sites.
The total free energy for the surface atoms is
    \begin{eqnarray}
        \label{eqn:freeEnergySurface}
        F_{\rm s}(n_{\rm s};N_{\rm s},T) &=& E^{\rm f}_{\rm s}(n_{\rm s}) + n_{\rm s} F^{\rm vib}_{\rm s}   \nonumber\\
            &&  \quad - k_B T \log \left( \frac{N_{\rm s}!}{(N_{\rm s}-n_{\rm s})! n_{\rm s}! } \right),
    \end{eqnarray}
with the vibrational component of the free energy, $F^{\rm vib}_{\rm s}$, taking the same form as that for the lattice gas, but potentially with a different frequency $\omega_{\rm s}$. Note that for this work, we have not explicitly calculated vibration frequencies for surface atoms, and instead set all vibration frequencies to be equal to that in the lattice.
The free energy of gas atoms in the surface, eqn \ref{eqn:freeEnergySurface}, can now be balanced against the free energy of the lattice gas, eqn \ref{eqn:freeEnergyLatticeGas}, and the free energy of the molecular hydrogen in the bubble, eqn \ref{eqn:GibbsFreeEnergyDiatomicGas}.

%Unfortunately, this leads to a transcendental equation, but the solution we seek is the number of surface atoms $n_s$ which satisfies
%    \begin{equation}
%        \mu_H = F^f_{\rm L} - F^b_s\left( n_s \right) + k_B T \log \left(\frac{n_s}{ N_s - n_s}\right).
%    \end{equation}

% At zero temperature, the solution is simpler,
%     \begin{equation}
%         n_s(T=0) = \frac{1}{2} \left( 
%             \sqrt{ \frac{4 N_s^2 (\alpha-E^f_L+\mu_H) }{\alpha \beta^2} + \frac{1}{3} }
%             - 1
%         \right)
%     \end{equation}

\subsection{Parameterization}

The fitting of the parameters to molecular statics calculations is described in \cite{Tirumala_JPCM2026}.
The values we use for bubbles in tungsten are given in table \ref{tab:parameters}.
$\alpha$, the maximum surface binding energy is fitted to the binding energy of ${\rm H}$ atoms to the void surface in the limit of large void size and small ${\rm H}$ atom count.
$\beta$, which defines the number of ${\rm H}$ atoms per surface area, is fitted to the monovacancy binding energy.
$\varepsilon$, the steric energy penalty, is fitted to zero temperature calculations of the energy of dense ${\rm {\rm H}_2}$ molecules in a box.
$\delta r$, the surface layer thickness, is fitted to exclude ${\rm H}_2$ molecules in small vacancy clusters in the low temperature limit.
\begin{table}[]
    \centering
    \begin{tabular}{c|ccc}
        parameter   &   value   & units    \\
        \hline
        $\alpha$                &   1.63    & eV           \\
        $\beta$                 &   5.4     &             \\
        $a_0$                   &   3.145   &   \r{A}        \\
        $\varepsilon$           &   1.22/$\Omega_0^4$   &   eV    \\
        $\delta r$              &       0.4 &  \r{A}           \\
        $E^{\rm f}_{\rm L}$     &   0.798   &   eV          \\
        $\omega_{\rm L}$        &   0.254   &   PHz         \\
        $\omega_{\rm s}$        &   0.254   &   PHz         
    \end{tabular}
    \caption{Parameters used to model thermodynamics of the hydrogen gas in a bubble, reproduced from ref \cite{Tirumala_JPCM2026}. Note $\Omega_0 = a_0^3/2$ is the volume per atom.}
    \label{tab:parameters}
\end{table}

\section{Convergence of lattice kMC study - H in vacancies }
\label{sec:convergence_soas_H_in_v}

\begin{figure*}
    \centering
    \includegraphics[width=0.7\linewidth]{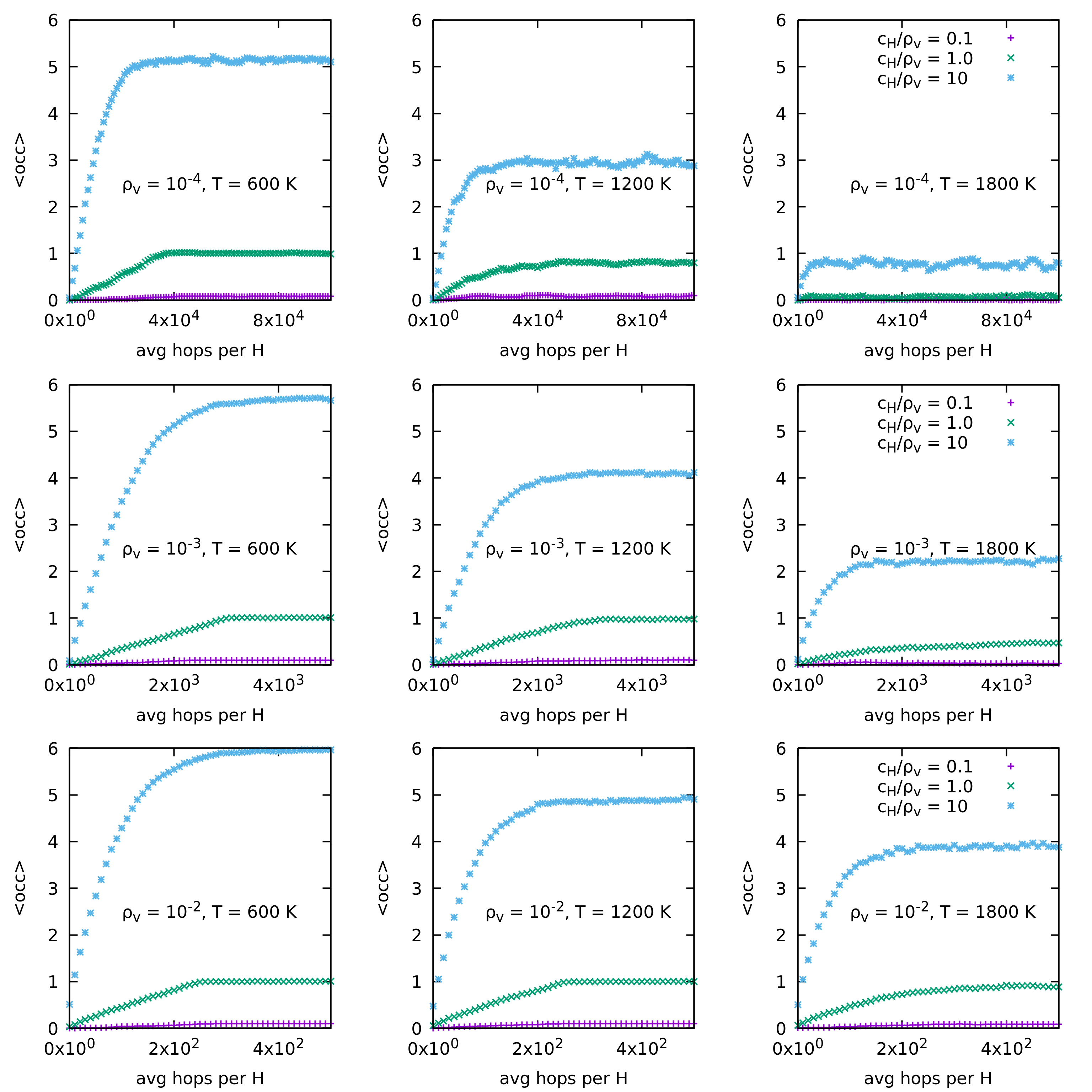}
    \caption{Convergence study for lattice kMC study of H in homogeneous vacancies in W.}
    \label{fig:convergence_H_in_vac}
\end{figure*}

\noindent
In this appendix, we present the convergence study proving that the dynamic steady state is rapidly reached for the vacancy concentration ranges used in this work.
We plot the occupation of H atoms per vacancy as a function of the number of H atom hops in figure \ref{fig:convergence_H_in_vac}.
Plotting other quantities such as the variance of H atoms per vacancy gives similar shaped curves.
Here we used the parameterization using the MNL2023 empirical interatomic potential.
The main factor influencing convergence rate was found to be the vacancy concentration, so here we plot three decades of vacancy concentration, $\rho_{\rm v} = 0.01,0.1,1.0 \%$ at.fr. 
The system size was chosen so that the number of H atoms total would not be excessive, here we used cubic supercells with $100,64,30$ unit cells per side for the three vacancy concentrations respectively, corresponding to 2M,520k,54k tungsten atom sites per supercell.
The H atom concentration affects the steady state occupation per vacancy reached, but does not greatly affect the convergence rate. 
The temperature strongly affects the time per hop, but only weakly affects the number of hops needed to reach steady state. Here we plot three decades of H concentration, $c_{\rm H}/\rho_{\rm v} = 0.1,1.0,10$.
We find that the steady state is reached in order 4/$\rho_{\rm v}$ hops per H atom. This scaling can be understood as reflecting the rate law for capture of a 3 dimensional walker, $r \sim 4 \pi R D_{\rm H}/\rho_{\rm v}$, where $R$ is a capture radius.
We conclude from this study that sampling the kMC simulation after 10000 hops per H atom gives converged answers for the concentrations considered, given that as the temperature is slowly reduced, the system remains close to steady state.
\\

\section{Binding energies of hydrogen atoms to vacancy clusters}
\label {sec:appendix_binding_energy}

\begin{figure*}
    \centering
    \includegraphics[width=0.7\linewidth]{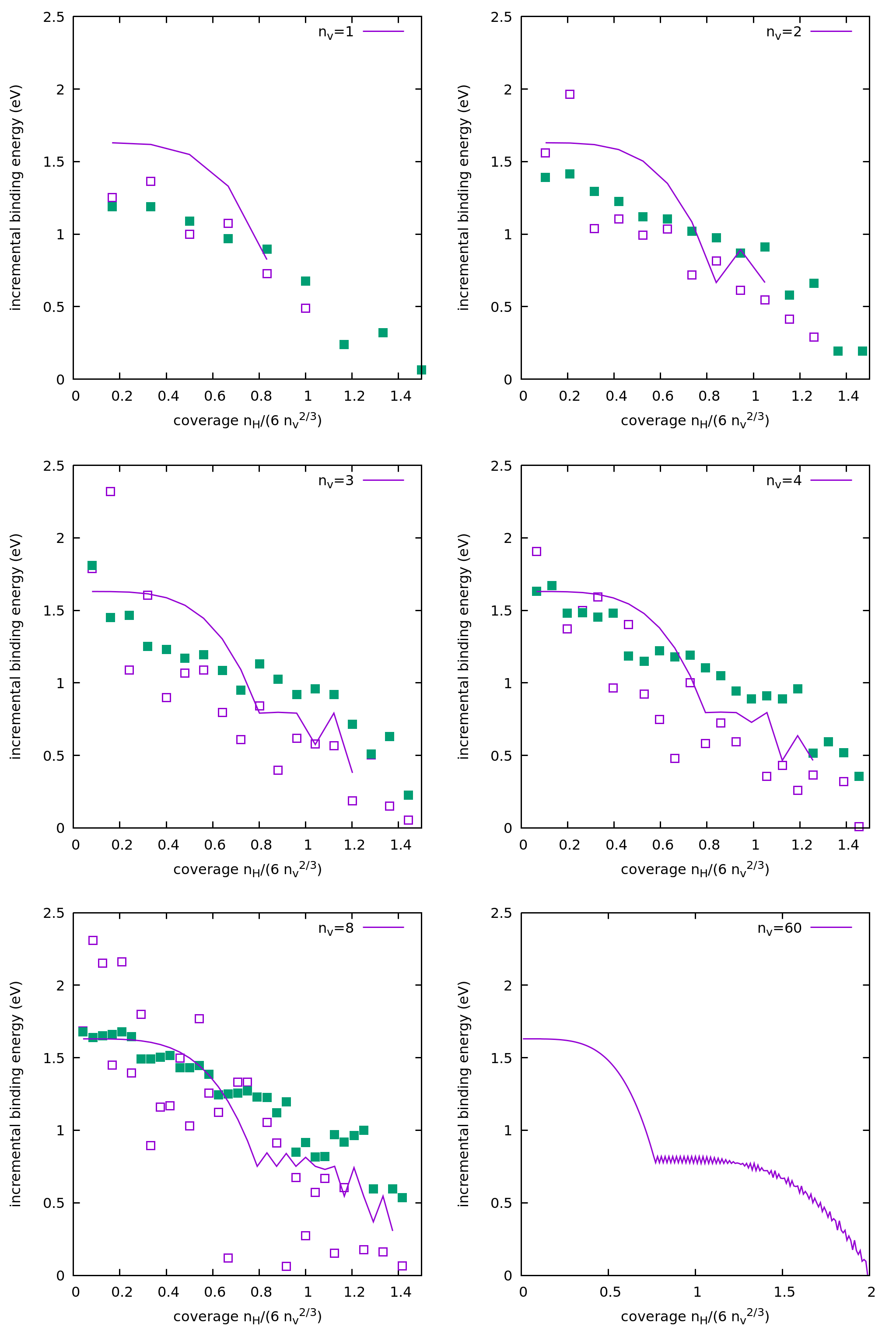}
    \caption{Computed incremental binding energies, with reference to the tetrahedral interstitial H atom, no zero point energy added. Open symbols--EAM potential used in this work~\cite{Mason_JPCM2023}. Solid points--density functional calculations~\cite{Hou_NatMat2019}. Solid line--empirical binding energy~\cite{Tirumala_JPCM2026}.
    }
    \label{fig:hydrogen_vacancy_binding_energies}
\end{figure*}

In this appendix, we report incremental binding energies used to make a comparison between the steady state solution (equation \ref{eqn:steady_state_solution}) using explicitly calculated binding energies and a simple temperature-dependent model suggested by ref \cite{Tirumala_JPCM2026}.
To generate the explicit binding energies, we set up a simulation cell of $7 \times 7 \times 7$ conventional bcc unit cells, and removed atoms in the centre to form the lowest energy vacancy cluster structure. Then we placed H atoms at random in the vicinity of the void, using positions corresponding to the tetrahedral and octahedral interstitial sublattices, and relaxed using conjugate gradients with {\texttt LAMMPS} using zero pressure boundary conditions and the MNL2023 tungsten-hydrogen potential~\cite{Mason_JPCM2023}.
We repeated the random placements up to 1000 times, to find a low energy configuration. Note that we cannot claim to have found the true minimum energy configuration by this method. 
The results are shown in figure \ref{fig:hydrogen_vacancy_binding_energies}. On this plot we also show the DFT calculations of Hou et al ~\cite{Hou_NatMat2019}.
Figure \ref{fig:hydrogen_vacancy_binding_energies} also shows the  zero temperature  empirical model of Tirumala et al~\cite{Tirumala_JPCM2026}. This model gives a jagged line despite the smooth underlying energy functions, as we constrain the system to have integer counts of surface atoms and molecules, and that the binding energy of both these types be positive. 
\\

\noindent
This plot demonstrates that the empirical potential may not be perfect, but has no significant errors  assuming DFT is the ground truth) for vacancy clusters that would lead to different qualitative conclusions in this work, and that the empirical model of Tirumala et al is also good, possibly (fortuitously) even better than the empirical potential on which it was based.
\\

\end{document}